\documentclass[a4paper,11pt]{article}
\pdfoutput=1 

\usepackage{jheppub} 

\usepackage[T1]{fontenc} 

\usepackage{xcolor}
\usepackage[utf8]{inputenc}
\usepackage{amsmath}
\usepackage{amssymb}
\usepackage{graphicx}
\usepackage{multirow}
\usepackage[normalem]{ulem}
\usepackage{comment}

\newcommand{\braket}[1]{\ensuremath{\left\langle #1 \right\rangle}}

\def\r {\rightarrow}
\def\l {\lambda}

\def\annone {\langle \sigma_{\rm ann,\,1}\, v\rangle}
\def\anntwo {\langle \sigma_{\rm ann,\,2}\, v\rangle}
\def\con {\langle \sigma_{\rm conv,\,12} \,v\rangle}

\title{\boldmath  Multi-component Dark Sectors: Symmetries, Asymmetries and Conversions
}

\author{Arnau Bas i Beneito\,,}
\author{Juan Herrero-Garc\'{\i}a}
\author{and Drona Vatsyayan}
\affiliation{Departamento de F\'isica Te\'orica, Universidad de Valencia and IFIC, Universidad de Valencia-CSIC,
C/ Catedr\'atico Jos\'e Beltr\'an, 2 | E-46980 Paterna, Spain}

\emailAdd{arnau.bas@ific.uv.es}
\emailAdd{juan.herrero@ific.uv.es}
\emailAdd{drona.vatsyayan@ific.uv.es}

\abstract{We study the relic abundance of several stable particles from a generic dark sector, including the possible presence of dark asymmetries. After discussing the different possibilities for stabilising multi-component dark matter, we analyse the final relic abundance of the symmetric and asymmetric dark matter components, paying special attention to the role of the unavoidable conversions between dark matter states. We find an exponential dependence of the asymmetries of the heavier components on annihilations and conversions. We conclude that having similar symmetric and asymmetric components is a natural outcome in many scenarios of multi-component dark matter. This has novel phenomenological implications, which we briefly discuss.}

\begin{document} 
\maketitle
\flushbottom

\section{Introduction\label{sec:intro}}

Several observations indicate the existence of non-visible matter termed as Dark Matter (DM), that comprises roughly $25\%$ of the total energy budget of the universe, and is therefore almost five times more abundant than the visible matter~\cite{Planck:2018vyg}. Despite this significant contribution, hardly anything is known about its composition, nature or dynamics. Among existing efforts and speculations made regarding the particle nature of DM, Weakly Interacting Massive Particles (WIMPs) are one of the simplest and most popular particle DM candidates, contributing to the energy density once they undergo thermal freeze-out. On the other hand, Asymmetric Dark Matter (ADM) models are motivated by the close relation between DM and baryonic energy densities, with $\rho_{\rm DM} \sim 5\rho_ {\rm B}$~\cite{Kaplan:2009ag}. Given that the baryon density is set by the baryon asymmetry of the universe, $\eta_B \sim 6 \times 10^{-10}$, the DM abundance could stem from an asymmetry in the dark sector in a similar manner (see Refs.~\cite{Petraki:2013wwa,Zurek:2013wia} for a review).  In the case of WIMPs, the value of the abundance is set by the thermally averaged annihilation cross section, whereas for ADM it is set by the initial asymmetry and the DM mass. Moreover, as discussed in Ref.~\cite{Graesser:2011wi}, there is also the possibility of an intermediate regime between the two extremes discussed above, where the DM is partially asymmetric and there is an exponential dependence on the annihilation cross section. 

However, most of these scenarios assume that the entire dark sector is composed of just one DM state. Given the plethora of stable particles in the Standard Model (SM) (electrons, protons, nuclei, neutrinos, photons...), it is reasonable to think that the dark sector could be made up of more than one stable component. Efforts are being made towards the study of such multi-component DM models, where the dark sector is comprised by several stable particles. For example, in Refs.~\cite{Liu:2011aa,Belanger:2011ww,Aoki:2012ub} the influence of DM conversions (annihilation of one DM state into another) in the final relic abundance was emphasised. Similarly, semi-annihilations can affect the final abundance~\cite{DEramo:2010keq}. Finally, multi-component DM, with its many degrees of freedom, leads to a richer phenomenology (see for instance Ref.~\cite{Zurek:2008qg,Bhattacharya:2016ysw,Gutierrez-Luna:2021tmq}), and can help to relax the existing bounds from direct detection, indirect detection and collider searches. Another proposal for multi-component dark sectors is Dynamical Dark Matter~\cite{Dienes:2011ja}. In this framework, the dark states come in large numbers and are unstable, with lifetimes proportional to their individual abundance. For a recent review of their possible collective effects see Ref.~\cite{Dienes:2022zbh}. 

The numerous states and interactions with SM and among themselves in a multi-component DM setup allow us to treat it as a complex system, which exhibits a behaviour that can be characterised by different scalings: power law and exponential. Such scaling laws are common in Physics and in other branches of science. Understanding this behaviour of multi-component DM subject to different interactions (annihilations and conversions) and conditions (symmetric/asymmetric) is one of the main goals of this paper. For that purpose, we will extend the analysis performed for one component asymmetric WIMP in Ref.~\cite{Graesser:2011wi} to two or more components with conversions among them.

The paper is structured as follows. In Section~\ref{sec:discsym}, we discuss the symmetries and combinations that are required to stabilise multiple states in the dark sector. In Section~\ref{sec:asymmetry}, we briefly discuss how DM asymmetries in different components can be generated from a $B-L$ asymmetry. In Section~\ref{sec:multicomp}, we develop the multi-component framework for both symmetric and asymmetric components. First, we study the final relic abundance of multi-component DM taking into account the unavoidable DM conversions. Then, in Section~\ref{sec:numerics}, we parameterise the possible particle physics models using effective interactions. We discuss the phenomenological implications of these scenarios in Section~\ref{sec:pheno} and provide some conclusions in Section~\ref{sec:summary}.

\section{Dark Sector Stabilisation Symmetries\label{sec:discsym}}

In this section, we briefly discuss the simplest possible symmetries to have multi-component DM. In DM models, usually a discrete and/or continuous global symmetry is imposed to stabilise the DM candidate. As pointed in Ref.~\cite{Walker:2009en}, continuous gauge symmetries alone cannot stabilise the DM and need to be spontaneously broken. However, they can motivate the origin of discrete gauge symmetries as a remnant of spontaneously broken gauge symmetries \cite{Krauss:1988zc}, such as matter parity (and $R$-parity in supersymmetry) from gauged $B-L$~\cite{Martin:1992mq}. In this work, we consider a multi-component DM framework that can be stabilized by adopting different symmetries, with the presence of both symmetric and asymmetric components. For symmetric DM, it is more natural to have a Majorana fermion or real scalar, stabilised by discrete symmetries. On the other hand, for asymmetric DM, the candidate needs to be a Dirac fermion or a complex scalar, stabilised by continuous symmetries. 

Let us start with the case of discrete symmetries. Most popular DM models employ a single $Z_2$ symmetry which guarantees the stability of the lightest $Z_2$ odd particle, allowing only one DM candidate in the model. In order to have multi-component DM, we would require a larger discrete symmetry or consider products of more than one discrete symmetry. We first investigate the number of stable DM candidates that are allowed for $Z_{{\mathcal N}\,>2}$, in particular for ${\mathcal N}=3, 4$, and then for the product of two or more discrete symmetries. 

Under an abelian $Z_{\mathcal N}$ symmetry, a field transforms as 
\begin{equation}
\phi \rightarrow \omega^q\,\phi\,,\quad \omega = \exp\left({2\pi i}/{{\mathcal N}}\right)\,,
\end{equation}
where $q = 0,1,2,\ldots, {\mathcal N}-1$ are the $Z_{\mathcal N}$ charges. Depending on how the fields transform under a given charge, we can divide them into classes. Class 0 corresponding to $q=0$ (implying that the field is neutral under the given symmetry) is by default common to all $Z_{\mathcal N}$ symmetries. For example, consider the case of $Z_2$, where $q=0,1$ and the field transforms as $\phi \r \phi$ (even) and $\phi \r -\phi$ (odd) respectively. Thus, there are only two classes for $Z_2 : \{0,1\}$. The lightest $Z_2$ even particle will be from the SM and the odd, the DM candidate. 

In the case of $Z_3$, we can assign charges $q=0,1,2$ to the fields. However, the charge $q=2$ corresponding to $\omega^2$ is same as assigning $q=-1$ since $\omega^2= \omega^\ast$, implying that the particle with charge $q=2$ is nothing but the antiparticle of the particle with $q=1$. Therefore, in the case of $Z_3$, we also have only two classes $\{0,1\}$ and there is only one DM candidate, the lightest particle transforming under a non-trivial $Z_3$ charge~\cite{Agashe:2010gt}.  

For $Z_4$, $\omega^q =1,i,-1,-i$ for $q=0,1,2,3$ respectively. Here, note that $q=3$ is equivalent to $q=-1$, and they would thus be placed in the same class. So, the classes for $Z_4$ are $\{0,1,2\}$. The lightest particle of class 1 $(i,-i)$, let's say $\phi_1$ is always stable and a DM candidate. The lightest particle of class 2 $(-1)$, say $\phi_2$ can also be stable, provided $m_{\phi_2} <2m_ {\phi_1}$. Thus, $Z_4$ guarantees the stability of at least one DM candidate, but can also allow for a second DM candidate in case its decays are kinematically forbidden to the particles belonging to class 1~\cite{Belanger:2012vp}. For DM models with $Z_3$ and $Z_4$ symmetries, see Refs.~\cite{Ma:2007gq, Belanger:2012zr, DEramo:2010keq, Belanger:2021lwd}.

In general, if ${\mathcal N}$ is a prime number, the lightest particle transforming with a non-trivial $Z_{\mathcal N}$ charge is always stable, and a potential DM candidate. One can also have heavier particles with $Z_{\mathcal N}$ charges as DM if their decay modes are kinematically forbidden. If ${\mathcal N}$ is a composite number, such that ${\mathcal N} = p_1^{s_1}\,p_2^{s_2}\ldots\, p_k^{s_k}$, where $p_i$ is a prime number ($p_i \neq p_j$ for $i \neq j$) and $s_i$ is natural, $Z_{\mathcal N}$ can be decomposed as a product of smaller groups~\cite{Batell:2010bp, Lee:2008pc}
\begin{equation}
Z_{\mathcal N} \simeq Z_{p_1^{s_1}} \times Z_{p_2^{s_2}} \times \ldots \times Z_{p_k^{s_k}}\,,
\end{equation}
and there can be at most $s_1+s_2 +\ldots +s_k$ stable particles depending on the kinematic conditions. Therefore, in order to realize multi-component DM with just one imposed discrete symmetry $Z_{\mathcal N}$, we require ${\mathcal N} \geq 4$. For studies of multi-component scalar DM in larger $Z_{\mathcal N}$ groups, see Ref.~\cite{Yaguna:2019cvp}.

Now we turn our discussion to the product of two or more of the same discrete symmetries. The simplest and most commonly used symmetry is $Z_2 \times Z_2^{'}$~\cite{Boehm:2003ha,Cao:2007fy,Belanger:2011ww,Bhattacharya:2016ysw,Maity:2019hre,Aoki:2012ub}. Such a choice of symmetry allows the existence of at least two stable particles and at most three, subject to kinematics. For the case of an $n$ $Z_2$ product: $Z_2 \times Z_2^{'} \times \dots \times Z_2^{(n)'}$, there are at least $n$ stable DM particles and at most $2^n -1$ stable particles can be DM candidates. It is easy to generalise this for the case of $Z_3$ and $Z_4$
\begin{align}\label{eq:dsdm}
Z_{\mathcal N} \times Z_{\mathcal N}^{'} \times \dots \times Z_{\mathcal N}^{(n)'} \Rightarrow 
\begin{cases}
n {\rm~at~ least}\\
c^n -1 {\rm ~at~ most\,,}
\end{cases}
\end{align}
where $c$ denotes the number of classes corresponding to the discrete symmetry $Z_{\mathcal N}$. For ${\mathcal N}=3$, since there are only two classes, the number of DM candidates for $Z_3 \times Z_3^{'}$ is equal to that of $Z_2 \times Z_2^{'}$ (see Ref.~\cite{Bhattacharya:2017fid} for a model with a $Z_3 \times Z_3^{'}$ symmetry). On the other hand, in the case of $Z_4$ which has three classes, $Z_4 \times Z_4^{'}$ allows for at max $3^2 -1 = 8$ DM candidates. The `$-1$' term guarantees stability of the lightest particle in the SM. 

We can also consider products of different discrete symmetries, for example $Z_2 \times Z_3$ (see Ref.~\cite{Choi:2021yps}), and generalise Eq.~(\ref{eq:dsdm}) as
\begin{align}
Z_{{\mathcal N}_1}^{n_1} \times Z_{{\mathcal N}_2}^{n_2} \times \ldots \times Z_{{\mathcal N}_k}^{n_k} \Rightarrow
\begin{cases}
n_1 + n_2 + \ldots + n_k {\rm~at~ least}\\
\prod_{i=1}^k {c_i}^{n_i}-1{\rm ~at~ most\,,}
\end{cases}
\end{align}
where ${\mathcal N}_i$ corresponds to the discrete symmetry used and $n_i$ denotes its multiplicity in the product.

In the following, we will also be interested in the case of multi-component DM with states that can be asymmetric. In this case, the most natural framework is the stabilisation of the DM components due to continuous global symmetries, which may be accidental. The possibilities we will implicitly consider are: \\
\emph{i)} a single global continuous $U(1)$ symmetry:
\begin{equation}\label{eq:U1a}
U(1)\,,
\end{equation}
with the DM $\chi_i$ having charge $q_i$ such that no linear terms can be written in $\chi_i$ at least at the renormalizable level. There could be linear terms at the non-renormalizable level, possibly suppressed by Planck-scale physics and therefore sufficiently suppressed \cite{Mambrini:2015sia}.\\
\emph{(ii)} the direct product of several $U(1)$s, such that each particle only transforms under one of them and is also the lightest particle charged under it,
\begin{equation}\label{eq:U1b}
U_1(1) \times  U_i (1) \times \dots \times U_N(1) \,,
\end{equation}
with DM $i$ transforming with the global charge $\chi_i \sim (0,\ldots,i=1,\ldots, 0)$. For multi-component DM models stabilised by the product of global symmetries, see Refs.~\cite{Liu:2011aa,Belanger:2011ww}.

\section{Asymmetries in the Dark Sector \label{sec:asymmetry}}

In general, DM may have both symmetric and asymmetric components. One of the simplest options is that a dark asymmetry can be generated from an asymmetry in the visible sector~\cite{Kaplan:2009ag}. The idea is that an asymmetry in the SM sector is transferred to the DM sector (for instance, via a generalised $B-L$). Standard Model (DM) operators  $\mathcal{O}_{\rm SM}\, (\mathcal{O}_i)$  with $Q_{B-L}\,(-\,Q_{B-L})$
charges generate a mixed operator,
\begin{equation} \label{op}
\mathcal{L}_{B-L=0} = \sum_i \frac{1}{\Lambda_i^n}\,\mathcal{O}_i\, \mathcal{O}_{\rm SM}\,,
\end{equation}
which is active at high energies and transfers any initial $B-L$ asymmetry in the SM to the DM sector. For definiteness, let us consider the case of Dirac DM, $\chi_i$.  We define $n^{+(-)}_i$ as the number densities of the $\chi_i$ particle (anti-particle), with $n_i^+>n_i^-$ without loss of generality, and similarly for baryons. The baryon number density, $B \equiv n^+_b - n^-_b$, can be expressed in terms of $B-L$ as $B \equiv b\, (B-L)$, where $b=30/97\, (28/79)=0.31\,(0.35)$ for $T < T_{\rm EW}\, (T > T_{\rm EW})$~\cite{Feng:2012jn}. The DM number density is $X_i \equiv n^+_i - n^-_i$, which can also be expressed in terms of $B-L$ as $X_i \equiv x_i\, (B-L)$. It is useful to express the number densities in terms of the yields $Y_i \equiv n_i/s$, where $s$ is the entropy density given by $s(T) =(2\pi^2/45)\, h_{\rm eff}(T) T^3$ and $h_{\rm eff}(T)$ are the effective degrees of freedom related to entropy. We also define $\eta_i = Y_i^+ - Y_i^-$, $\eta_B = Y_b^+ - Y_b^-$, and their ratio
\begin{equation} \label{eq:eps}
\epsilon_i \equiv\frac{x_i}{b} \equiv \frac{\eta_i}{\eta_B} \,.
\end{equation}

Depending on whether the interaction rate generated by the transfer operator, Eq.~\eqref{op}, is active below or above the EW scale, one obtains~\cite{Harvey:1990qw,Feng:2012jn}:
\begin{itemize}
\item $T_{\rm int} <T_{\rm EW}$: $x_i=-11/111\, Q_{B-L}$, $\epsilon_i=-0.32\, Q_{B-L}$\,.
\item $T_{\rm int} >T_{\rm EW}$: $x_i=-11/79\, Q_{B-L}$, $\epsilon_i=-0.39\, Q_{B-L}$\,. 
\end{itemize}
If there are strong dark interactions (large annihilations), the symmetric component may be totally erased. In that case, in order to reproduce the DM abundance, we require
\begin{equation}\label{eq:cosco}
\frac{\Omega_{\rm DM}}{\Omega_{\rm B}}\simeq 5 =\frac{\sum_i \epsilon_i \,m_i}{m_b}\,.
\end{equation}
If this is the case, there is a prediction for the combination of DM masses. It only depends on the energy at which the operator is active, and on how the asymmetry is transferred, i.e., on the $B-L$ charge. For example, in the case of 2DM there are different possibilities:
\begin{enumerate}
\item Identical DM particles with same charges, e.g. $ B-L \rightarrow  X_1, X_2=X_1 $: the same operator couples to both DM particles. For instance, if the operator is active above (below) the EW scale, one obtains 
\begin{equation} \label{2DMequal}
m_1+m_2 \simeq -\frac{13\, (16)}{Q_{B-L}}\,\rm{GeV}\,.
\end{equation}
In the case of $N$ components, we get
\begin{equation} \label{NDMequa}
\sum_{i=1}^{N} m_i= -\frac{13\, (16)}{Q_{B-L}}\,\rm{GeV}\,.
\end{equation}
Therefore, the larger the number of asymmetric states, the lower their masses. Indeed, if all masses are similar, the presence of $N$-asymmetric components would translate into an $N$-suppressed mass-scale compared to the 1-DM scenario,
\begin{equation} \label{NDMequal}
m^{\rm N-DM} \simeq \frac{m^{\rm 1-DM}}{N}\,.
\end{equation}
Notice that this can be understood as an upper limit on the mass scale in order for the DM not to be overabundant. Of course, as for 1DM, heavier DM is allowed if the DM asymmetry is unrelated or suppressed compared to the baryonic one.

\item Different DM particles with different charges, e.g., $B-L \rightarrow  X_1, X_2\neq X_1$: different operators couple to both DM particles. For example, if operator one (two) with $B-L$ charge $Q_1$ ($Q_2$) is active above (below) the EW scale, one gets
\begin{equation}\label{eq:adm2}
\frac{m_1\,Q_1}{6.5\,\rm{GeV}}+\frac{m_2\,Q_2}{8\,\rm{GeV}} \simeq 2\,.
\end{equation}
In the case of $N=(N_1+N_2)$ DM components, where $N_1~(N_2)$ components are coupled to operators active above (below) the EW scale, the above equation generalizes to
\begin{equation}  \label{eq:Ndm2}
\sum_{i=1}^{N_1} \frac{m_i Q_i}{(13\,{\rm GeV}/N)} + \sum_{j=1}^{N_2} \frac{m_j Q_j}{(8\,{\rm GeV}/N)} = N\,.
\end{equation}
Note how Eq.~\ref{2DMequal} (Eq.~\ref{NDMequa}) is a particular case of Eq.~\ref{eq:adm2} (Eq.~\ref{eq:Ndm2})  for identical charges above (below) the EW scale.

\item Different DM particles with different charges, $B-L \rightarrow  X_1 \rightarrow  X_2$. Only one of them, $\chi_1$, is coupled to the SM; the asymmetry is first transferred to one DM state and then there is a further sharing of the asymmetry with another DM state. For example, this can be generated by the operators
\begin{equation}
\mathcal{L}_{B-L=0} = \frac{1}{M^3_1}\,\chi^3_1 L H +\frac{1}{M_2}\,\chi^2_1 \phi^2_2\,,
\end{equation}
with $\phi$ being a scalar DM particle. This implies $x_2 = 2\,x_1$ and therefore\footnote{The generalisation to $N-$component DM is very model-dependent and should be further investigated.}
\begin{equation}
\frac{m_1}{6.5\,\rm{GeV}}+\frac{m_2}{3.25\,\rm{GeV}} = 2\,.
\end{equation}

\end{enumerate}
In the next section we consider the general case where both asymmetric and symmetric components may be present. If the asymmetric component dominates, the previous mass relations are very good approximations.

\section{Multi-component Framework  \label{sec:multicomp}}

In the following, we set up the framework for multi-component DM with both symmetric and asymmetric components. Let us consider an $N$-component DM scenario stabilised by a certain choice of continuous $U(1)$ symmetries, as discussed in Section~\ref{sec:discsym}, with $m_1 > m_2 > \ldots > m_N >m_{\rm SM}$. Within this setup we may have one or more of the following reactions:
\begin{enumerate}
\item Annihilations $\chi_i \bar\chi_i \rightarrow {\rm SM \,\overline{SM}}$. We use the notation $\sigma_{\rm \chi_i \bar\chi_i
\rightarrow SM\,SM} \equiv  \sigma_{\rm ann,\,i}$.
\item Coannihilations $\chi_i\, \bar\chi_j \rightarrow {\rm SM \,\overline{SM}}$, where $\chi_j$ is a state close in mass to $\chi_i$. We can neglect coannihilations if $\Delta_{ij} \equiv (m_i-m_j)/m_j \gtrsim0.05$.
\item Conversions $\chi_i \bar\chi_i \rightarrow \chi_j \bar\chi_j$, for $m_i>m_j$, which change the individual DM number densities but not the total one \footnote{Note that the forbidden channel processes such as $\chi_j \bar{\chi_j} \r \chi_i \bar{\chi_i}$, i.e. annihilation of lighter DM to heavier DM are operative at high temperatures and may be significant for small splittings, $\Delta_{ij}<1$ \cite{Griest:1990kh,DAgnolo:2015ujb,Aoki:2016glu,Wojcik:2021xki,Yang:2022zlh}. The Boltzmann equations already take this effect into account.}. We use the notation $\sigma_{\rm \chi_i \bar\chi_i
\rightarrow \chi_j \bar\chi_j} \equiv  \sigma_{\rm conv,\, ij}$.
\item Semi-annihilations $\chi_i \chi_j \r \chi_k {\rm~ SM}$, for $N>2$. These processes depend on the symmetry imposed. In this case, DM needs to carry some SM charge.
\end{enumerate}

In the following, we consider only the case of annihilations and conversions. We are interested in conversions of the form $\chi_i \bar\chi_i \leftrightarrow \chi_j \bar\chi_j$, which are not forbidden by any symmetry that allows co/annihilations and are therefore typically always present. This is the case if both $\chi_i$ and $\chi_j$ are charged under $Z_2$ and $Z_4$ (class 2) symmetries respectively. Other conversions such as $\chi_i \chi_i \chi_i \leftrightarrow \chi_j \chi_j$, for example, are phase-space suppressed and allowed only if $\chi_i$ is charged under $Z_3$, and $\chi_j$ under $Z_2$. Similarly, conversions of the form $\chi_i \chi_i \chi_j \rightarrow \bar\chi_i \bar\chi_j \bar\chi_j$ are also phase-space suppressed. 

Furthermore, we always have annihilations that are significant and comparable to conversions. This avoids significant reheating in the dark sector if DM states become non-relativistic and their temperatures increase exponentially~\cite{Berlin:2016gtr}. The non-negligible coupling to the SM also ensures that they are thermalised with the SM plasma at freeze-out. Therefore, considering only $2 \rightarrow 2$ processes, assuming kinetic equilibrium with the SM of all species and neglecting the quantum statistics for the species involved, the evolution equation for the number densities of the particles read
\begin{align} \label{eq:Bn}
\frac{dn^\pm_i}{dt}+3Hn^\pm_i = &-{\langle \sigma_{\rm ann,\,i}\, v\rangle}\, (n^+_i n^-_i- \bar n_{i}^{+} \bar n_i^{-}) -\sum_{i<j}{\langle \sigma_{\rm conv, \,ij}\, v \rangle}\, (n^+_i n^-_i- \frac{\bar n_i^{+} \bar n_i^{-}}{\bar n_j^{+} \,\bar n_j^{-}}n_j^{+}n_j^{-})\nonumber\\
&+\sum_{i>j}{\langle \sigma_{\rm conv,\, ij}\, v \rangle}\, (n^+_j n^-_j- \frac{\bar n_j^{+} \bar n_j^{-}}{\bar n_i^{+} \,\bar n_i^{-}}n_i^{+}n_i^{-})\,,
\end{align}
where $\bar n_i^\pm = n_{\rm eq}e^{\pm \xi}$ denote equilibrium number densities of the species, for which we use Maxwell-Boltzmann distributions, $n_{\rm eq}=g(mT/2\pi)^{3/2}\, e^{-m/T}$, with $\mu=\xi T$ the chemical potential and $g$ the number of internal degrees of freedom. Throughout this work, we denote by $\langle \sigma_{\rm ann,\,i}\, v\rangle$ and $\langle \sigma_{\rm conv,\, ij}\, v \rangle$ the thermally-averaged annihilation and conversion cross sections, respectively. The Hubble parameter is given in the radiation-dominated epoch by $H (T) = (8\pi^3/90)^{1/2 }\,g^{1/2}_{\rm eff}(T)\, T^2/M_{\rm Pl}$, with Planck's mass  $M_{\rm Pl}$ equal to $1.2\times 10^{19}$ GeV, and $g_{\rm eff}(T)$ the effective relativistic degrees of freedom.\footnote{We take the Hubble parameter to be dominated by the SM degrees of freedom by assuming that there are no significant numbers of relativistic degrees of freedom in the dark sector.} Furthermore, the temperature evolution of the Universe can be expressed by scaling it with the mass of the heaviest component ($m_1$ in our case) using $x=m_1/T$, so that Eq.~\eqref{eq:Bn} can be written in terms of the yields as
\begin{align}\label{eq:By}
\frac{dY^\pm_i}{dx}= &-\frac{s}{H x}\Bigg[{\langle \sigma_{\rm ann,\,i}\, v\rangle}\, (Y^+_i Y^-_i- \bar Y_{i}^{+} \bar Y_i^{-}) -\sum_{i<j}{\langle \sigma_{\rm conv,\, ij}\, v_\rangle}\, \left(Y^+_i Y^-_i- \frac{\bar Y_i^{+} \bar Y_i^{-}}{\bar Y_j^{+} \,\bar Y_j^{-}}Y_j^{+}Y_j^{-}\right)\nonumber\\
&+\sum_{i>j}{\langle \sigma_{\rm conv,\, ij}\, v_\rangle}\, \left(Y^+_j Y^-_j- \frac{\bar Y_j^{+} \bar Y_j^{-}}{\bar Y_i^{+} \,\bar Y_i^{-}}Y_i^{+}Y_i^{-}\right)\Bigg]\,.
\end{align}
For ease of analysis and to highlight the qualitative aspects of the setup, we first consider a two-component DM system and discuss the cases where both components are symmetric or asymmetric. Later on, we extend the analysis to more complex sectors.

\subsection{Symmetric components \label{sym_conv}}
For 2DM with symmetric components, Eq.~\eqref{eq:By} can be written as
\begin{align}\label{eq:sym}
\frac{dY_1}{dx}&=-\frac{s}{H x}\bigg[\annone \,(Y_1^2 - \bar Y_1^2 ) 
+\con (Y_1^2 -\frac{Y_2^2}{\bar Y_2^2}\bar Y_1^2)\bigg]\,, \nonumber\\
\frac{dY_2}{dx}&=-\frac{s}{Hx}\bigg[\anntwo \,(Y_2^2 - \bar Y_2^2 ) 
-\con (Y_1^2-\frac{Y_2^2}{\bar Y_2^2}\bar Y_1^2)\bigg]\,.
\end{align}

If conversions are neglected, the Boltzmann Equations in Eqs.~\eqref{eq:sym} decouple, and in order to reproduce the observed relic abundance, we require (for $m_i \geq \mathcal{O}({\rm GeV})$ and assuming $s$-wave annihilations),
\begin{equation}\label{eq:s1}
\Omega h^2 \sim \frac{1}{\langle \sigma_{\rm ann,\,1}\, v\rangle}+\frac{1}{\langle \sigma_{\rm ann,\,2}\, v\rangle} \equiv \frac{1}{\langle \sigma v \rangle}_{\rm eff} \simeq \frac{1}{2.2\cdot 10^{-26} \,\rm{cm}^3/\rm{s}}= 1.37 {\rm~pb}^{-1}\,.
\end{equation}
Therefore, the abundance is dominated by the particle with the \emph{smallest} annihilation cross section, i.e., it is like 1DM, unless $\langle \sigma_{\rm ann,\,1}\, v_1\rangle \simeq \langle \sigma_{\rm ann,\,2}\, v_2\rangle$. If both particles have similar annihilation cross sections, the yield is dominated by the lightest particle, and the contributions to the relic abundance from both states are of similar size.

\begin{figure}[!htb]
\centering
\includegraphics[scale=0.4]{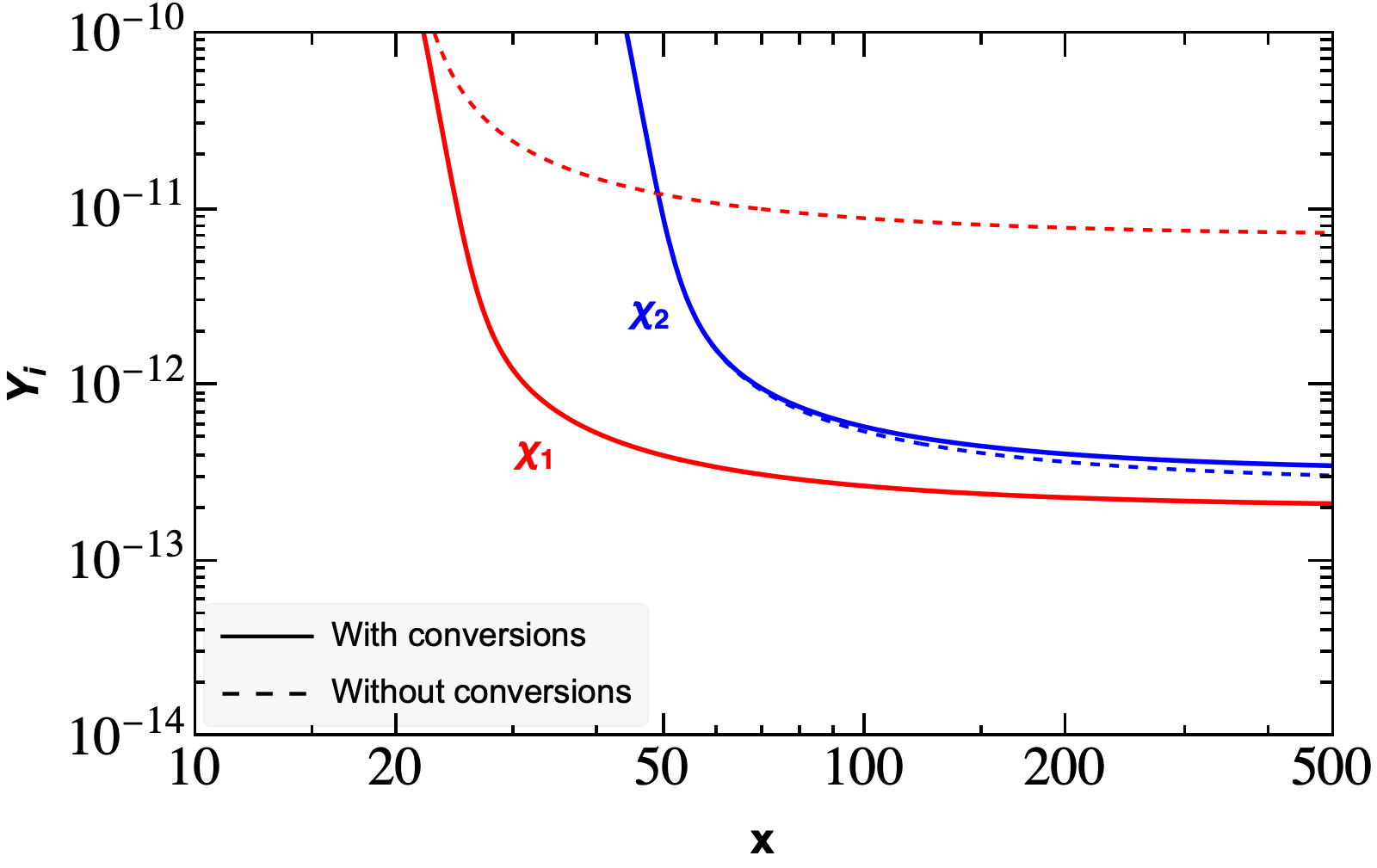}
\caption{Evolution of yields $Y_{1,2}$ (solid curves) as a function of $x=m_1/T$ with $\anntwo=6 \times 10^{-9}{\rm~GeV}^{-2},\annone=0.1 \times 10^{-9}{\rm~GeV}^{-2},\con=4\times 10^{-9}{\rm~GeV}^{-2}$ for $m_1=1000 {\rm~GeV}$ and $m_2=500 {\rm~GeV}$. The dashed curves correspond to the case without conversions i.e. $\con = 0$. Notice that the freeze-out of $\chi_2$ happens at $x = m_1/T \simeq 40 \Rightarrow m_2/T  \simeq  20$. A similar result was obtained in Ref.~\cite{Aoki:2012ub}.}
\label{fig:symcon}
\end{figure}

On the other hand, if conversions are significant, both components may end up with similar abundances, i.e., DM may be multi-component. For instance, assuming the hierarchy ${\langle \sigma_{\rm ann,\,2} \,v\rangle} 
\gg{\langle \sigma_{\rm conv} \,v\rangle}
\gg{\langle \sigma_{\rm ann,\,1} \,v\rangle}$, the abundance of $\chi_1$ ($\chi_2$) gets reduced (increased) with respect to the case without conversions~\cite{Liu:2011aa,Belanger:2011ww,Aoki:2012ub,Bhattacharya:2016ysw, Maity:2019hre}. This behaviour is illustrated in Fig.~\ref{fig:symcon}, where we show the evolution of abundances for $\chi_1$ and $\chi_2$ for the assumed hierarchy of cross sections in presence and absence of conversions. With conversions, Eq.~\eqref{eq:s1} is roughly modified as
\begin{equation}\label{eq:s2}
\Omega h^2 \sim \frac{1}{\annone + \con}+\frac{1}{\langle \sigma_{\rm ann,\,2}\, v\rangle} \simeq \frac{1}{\langle \sigma v \rangle}_{\rm eff} \,.
\end{equation}   
It is clear from this that conversions reduce the total relic abundance, and so, in comparison to the single-DM-component case, smaller annihilations into SM states are necessary. This is turn allows to have smaller couplings between the visible and the dark sector, and therefore, to relax existing constraints.

We can extend the above discussion to $N$ components. Let us consider the case where there are $N$ DM components that have similar cross-sections for annihilations and conversions, e.g. $\braket{\sigma_{\rm ann,i}v} = \braket{\sigma_{\rm conv,ij}} = \braket{\sigma v}$ for all $i,j$. This simplifies the analysis so that the only parameters of interest are the individual DM masses $m_i$ and $\braket{\sigma v}$. For example, let us consider the case of $N=10$ and take $m_i - m_{i+1}=100 {\rm~GeV}$, with $m_1=1000 {\rm~GeV}$. 

\begin{figure}[!htb]
\centering
\includegraphics[width=.49\linewidth]{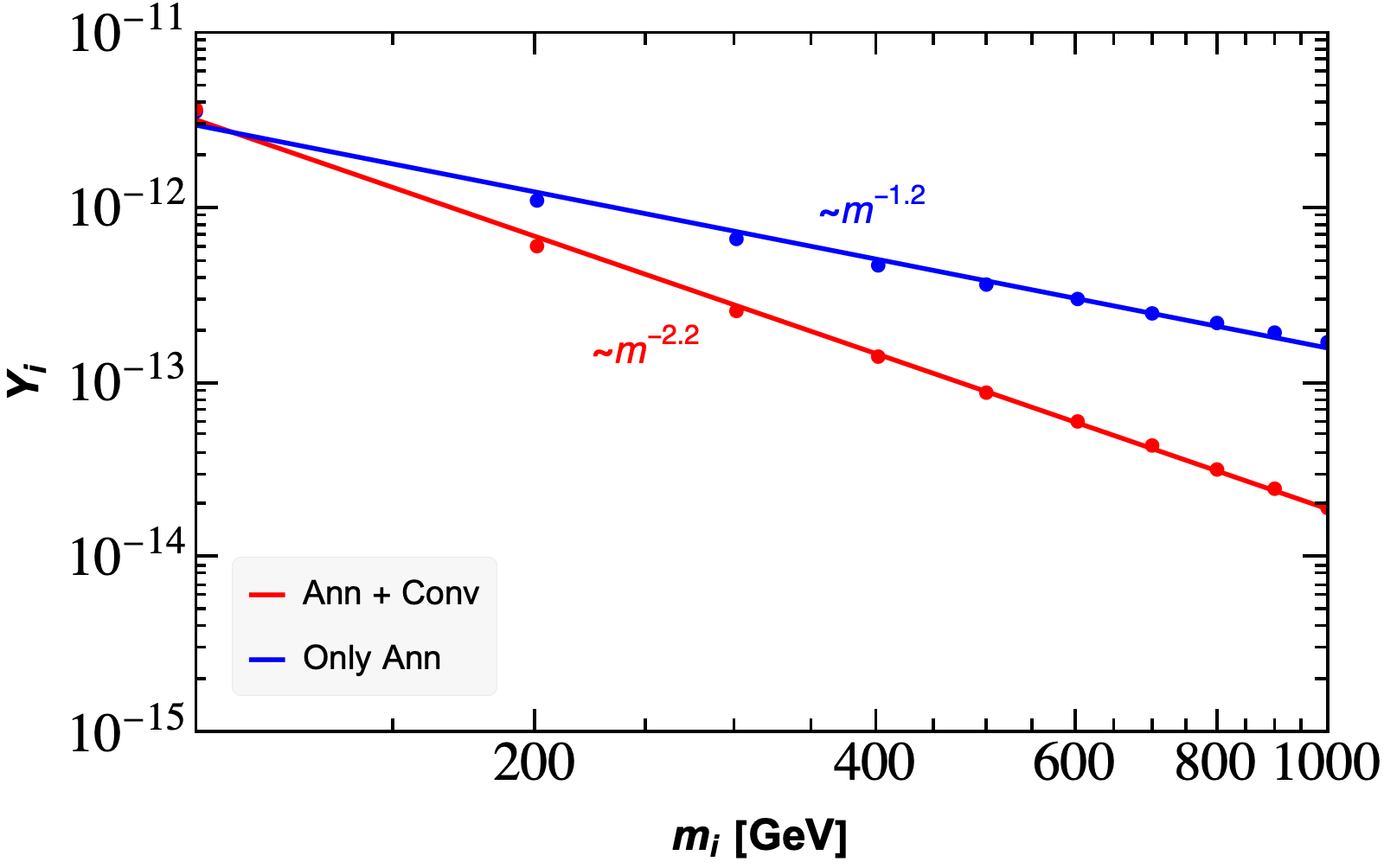}
\includegraphics[width=.49\linewidth]{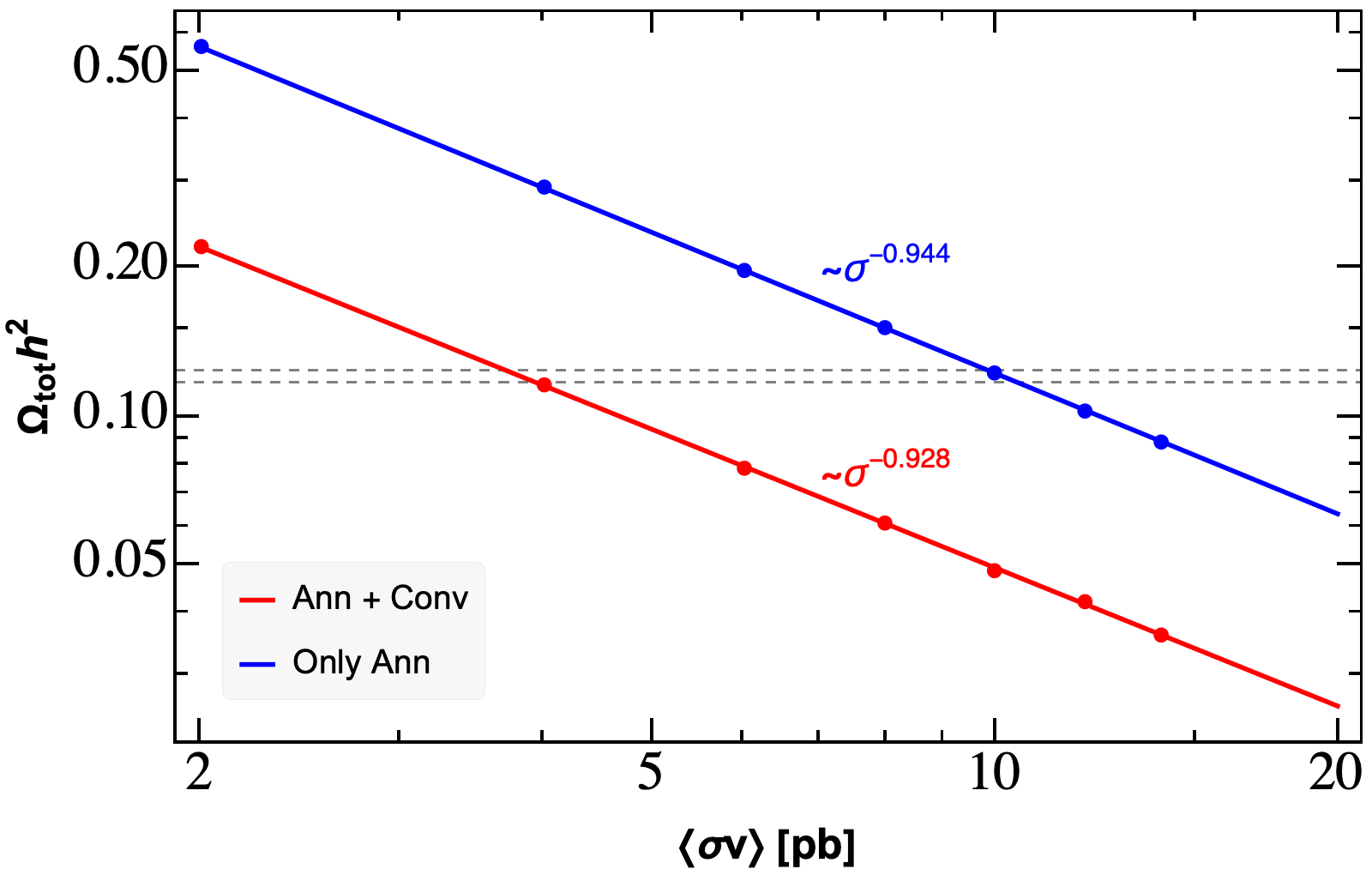}
\caption{Power-law fits for the yields (left panel) and total abundance (right panel) in the symmetric case for 10 DM particles with equal annihilation and conversion cross section, $\braket{\sigma v}$. \emph{Left panel:} Individual yields versus DM mass; we choose  $\braket{\sigma v} =2{\rm~pb}$. \emph{Right panel:} Total abundance vs cross section; the grey dashed lines represent the $3\sigma$ range for the observed DM relic abundance, $\Omega_{\rm DM}h^2=0.12 \pm 0.0012$~\cite{Planck:2018vyg}.}
\label{fig:y10}
\end{figure}  
 
In Fig.~\ref{fig:y10}, we plot the individual asymptotic yields (left panel) and total abundance (right panel), as well as the fits to power laws of DM mass. Since the cross section is equal for all components, the yield goes as $\sim m^{-1}$ for the case without conversions and $\sim m^{-2}$ for the case with conversions. The inverse dependence on mass is well known from the analytic expression for the yield in the standard freeze-out case without conversions, as all the Boltzmann Equations are decoupled. The additional mass suppression in the case of conversions can be understood as follows: the heavier the mass of the DM component, the larger the number of channels for conversions. For example, $\chi_1$, being the heaviest component, can undergo conversions to all the other 9 components whereas for $\chi_9$, the only allowed conversion is $\chi_9 \chi_9 \r \chi_{10}\chi_{10}$. This implies that the individual abundances are almost independent of mass in the case without conversions, whereas in presence of conversions, they are inversely proportional to mass, as $\Omega_i \propto m_i Y_i$. For the lightest component, the yields are identical in both cases since there are no channels for its conversion and the excess due to its production from all the other components gets annihilated to SM. In Fig.~\ref{fig:y10} (right panel), we can see that the total abundance is inversely proportional to the cross section in both the cases. Furthermore, the presence of conversions allows us to reproduce the correct abundance for smaller values of the annihilation cross-section (roughly a factor of 2), and thus in general of the DM couplings to SM. This is a generic feature, which as argued before translates in a relaxation of the experimental bounds.

Finally, let us mention that for cold thermal relics, there exists a bound on how heavy the DM particle can be from unitarity in order not to overclose the universe: $m_{\chi} \lesssim 110{\rm~TeV}$~\cite{Griest:1989wd, Baldes:2017gzw}. In the case of multi-component DM, this translates into a bound on the masses of the DM components for s-wave annihilations,
\begin{equation}
\left(\sum_i^N m_i^2\right)^{1/2} \lesssim 110{\rm~TeV}\nonumber\,,
\end{equation}  
when only annihilations are considered. When conversions are present, the bound reads\footnote{Here, we assume $s$-wave approximation for both conversions and annihilations.} 
\begin{equation}
\left(\sum_i^N \frac{m_i^2}{(N-i+1)}\right)^{1/2} \lesssim 110{\rm~TeV}\,.\nonumber
\end{equation}
Therefore, as for the case of asymmetric DM, the existence of a complex dark sector points to lower DM masses. Again, if all masses are similar, the presence of $N$-symmetric components translates into an $\sqrt{N}$-suppressed mass-scale compared to the 1-DM scenario,
\begin{equation} \label{NDMequalsym}
m_{\rm sym}^{\rm N-DM} \lesssim \frac{m_{\rm sym}^{\rm 1-DM}}{\sqrt{N}} \lesssim  \frac{110{\rm~TeV}}{\sqrt{N}}\,,
\end{equation}
where in the last step we assume $s$-wave annihilations and conversions. Notice how, now, the suppression with $N$ is milder as compared to the asymmetric case, c.f. Eq.~\ref{NDMequal}. The above bounds may be relaxed to $190 {\rm~TeV}$ ($220{\rm~TeV}$) for $p$-wave (for a combination of $s$- and $p$-wave) annihilations and conversions \cite{Baldes:2017gzw}.

\subsection{Asymmetric components}

\subsubsection{Analytical expressions} \label{sec:an}

Now we study how the picture changes if there are initial DM asymmetries. We follow closely the work of Ref.~\cite{Graesser:2011wi} and extend it to the case of 2DM with initial asymmetries $\eta_i = Y_i^+ - Y_i^-$, $i=1,2$. Of course, the asymmetries remain constant and equal to the initial value throughout the evolution. Following their notation, we can define the asymmetric ratios $r_i \equiv Y_i^-/Y^+_i$, which are in the range $0\leq r_i \leq 1$, where the upper (lower) limit is for totally (a)symmetric DM. In this way, the co-moving number densities $Y^\pm_i$ can be completely expressed in terms of $\eta_i$ and $r_i$ as 
\begin{equation}\label{eq:ypym}
Y^+_i=\eta_i/(1-r_i)\,,{\rm~and~} Y^-_i=r_i Y^+_i\,.
\end{equation}

Given that the chemical potentials of particles and their antiparticles are equal and opposite, we define $\bar{r_i} \equiv e^{-2\xi(x)}$ with $\xi=\sinh^{-1}(\eta_i/(2\bar Y_{i}))$. Hence, Eq.~\eqref{eq:By} can be written for the asymmetric ratios $r_i$ in the suggestive forms
\begin{align} \label{eq:B2}
\frac{dr_1}{dx} = &-\frac{s \eta_1}{H x}\bigg[\annone \left(r_1-\bar r_{1}\zeta^2(r_1)\right) +\con \left(r_1-\frac{r_{2}\bar r_{1}}{\bar r_{2}} \frac{\zeta^2(r_1)}{\zeta^2(r_2)}\right)\bigg]\,,\nonumber\\
 \frac{dr_2}{dx} = &-\frac{s \eta_2}{H x}\bigg[\anntwo\left(r_2-\bar r_{2}\zeta^2(r_2)\right)-\con \frac{\eta_1^2 (1-r_2)^2}{\eta_2^2 (1-r_1)^2}\left(r_1-\frac{r_{2}\bar r_{1}}{\bar r_{2}}\frac{\zeta^2(r_1)}{\zeta^2(r_2)}\right)\bigg]
 \nonumber\,,\\
\end{align}
where $\zeta(r_i) \equiv {(1-r_i)}/{(1-\bar r_{i})}$. Notice that if the initial asymmetries $\eta_i$ are zero, the ratios $r_i$ would remain constant and equal to one, like in the standard freeze-out case. 

It is interesting to derive some analytical solutions to better understand the behavior of the system. Considering the evolution of $r_i$ ($i=1,2$), we have two different regimes:
\begin{itemize}
\item For $x<x_{f_i}$, there is thermal equilibrium and $r_i\,(x) \simeq \bar r_i\,(x)$.

\item For $x>x_{f_i}$, $r_i\,(x) \gg \bar r_i\,(x)$ and also $\bar r_1(x)/\bar r_2(x)\ll 1$, 
\end{itemize}
where $x_{f_i}$ denotes the freeze-out temperature for the species. We can expand the annihilation and conversion cross sections in the relative velocity and express the thermally averaged cross sections as\footnote{Note that we expand the conversion cross section in terms of the velocity of the heavier component.} $\langle \sigma_{\rm ann,i}\, v \rangle\equiv \sigma_i\,(T/m_i)^{k_i}$ and  $\langle \sigma_{\rm conv} \,v_1\rangle\equiv \sigma_c\,(T/m_1)^{k_c}$, where one can consider different combinations of $s$-wave ($k_{i/c}=0$) and $p$-wave ($k_{i/c}=1$) cross sections.\footnote{In the numerical analysis we consider an $s$-wave annihilation cross section. Similar results can be obtained for higher partial waves.} We can write
\begin{align}
\frac{s}{H x}\braket{\sigma_{\rm ann,i}v}&=\lambda_{\rm a,i}\,g^{1/2}_* x^{-k_i -2}\left(\frac{m_1}{m_i}\right)^{k_i}\,,\nonumber\\
\frac{s}{H x}\braket{\sigma_{\rm conv}v}&=\lambda_{\rm a,1}\,f_c\, g^{1/2}_* x^{-k_c -2}\,,
\end{align}
where $\lambda_{\rm a,i}=(\pi/45)^{1/2}\, M_{\rm Pl}\, m_1\, \sigma_i$, $f_c\equiv \sigma_c/\sigma_1$, and 
\begin{equation}
g^{1/2}_*(x) = \frac{h_{\rm eff}}{g_{\rm eff}^{1/2}}\left(1- \frac{1}{4} \frac{x}{g_{\rm eff}} \frac{dg_{\rm eff}}{dx}\right)\,.
\end{equation}

In the following, for simplicity, we consider the case: $k_1 = k_2 = k_c \equiv k$, $\eta_1 = \eta_2 \equiv \eta$, $\lambda_{\rm a,2} \simeq \lambda_{\rm a,1} \equiv \lambda_{\rm a}$ and approximate $\zeta(r_i) \sim 1$. Therefore, with the above approximations, we obtain
\begin{align}\label{eq:r1r2}
\frac{dr_1}{dx} \simeq &-  \lambda_{\rm a}\,\eta\, g^{1/2}_*\, x^{-k-2} \,(1+f_c)r_1\,,
 \nonumber\\
 \frac{dr_2}{dx} \simeq &-  \lambda_{\rm a}\,\eta\, g^{1/2}_*\, x^{-k-2} \,(r_2 -f_c r_1)\,.
\end{align} 
Given that $m_2 < m_1$, for similar cross sections $r_2 \gg r_1$, and hence we can neglect $f_c r_1$ in the second line of Eq.~(\ref{eq:r1r2}) and solve the decoupled system
\begin{align}\label{eq:analytic}
r_1(x) &\simeq\,  \bar{r}_{1,f}\, e^{- \lambda_{\rm a}\,(1+f_c)\,\eta\,\Phi_1(x,m_1)}\,,
 \nonumber\\
r_2(x) &\simeq\, \bar{r}_{2,f}\, e^{- \lambda_{\rm a}\,\eta\,\Phi_2(x,m_2)}\,,
\end{align}
where $\bar{r}_{i,f} \equiv \bar{r}_i(x_{f_i})$ and
 \begin{equation}\label{eq:phi}
\Phi_i(x,m_i) \equiv \int^x_{x_{f_i}} dx^\prime x^{\prime -k-2} g_*^{1/2}\,.
 \end{equation}
Therefore, the DM asymmetry ratios decrease exponentially with the annihilation cross-section (i.e., DM becomes more asymmetric as annihilations are increased), in a similar way as for 1DM scenarios~\cite{Graesser:2011wi}. Furthermore, in presence of significant conversions, the asymmetry ratio for the heavier component also decreases exponentially with the conversion cross section. This is analogous to the symmetric case (see Eq.~\eqref{eq:s2}) where the yield for the heavier components was inversely proportional to the sum of annihilation and conversion cross sections whereas that for the lighter one it was only inversely proportional to its annihilation cross section. In this case, however, the crucial difference is that there is an exponential dependence. 

\subsubsection{Numerical analysis}
\begin{figure}[!htb]
\centering
\includegraphics[width=0.49\textwidth]{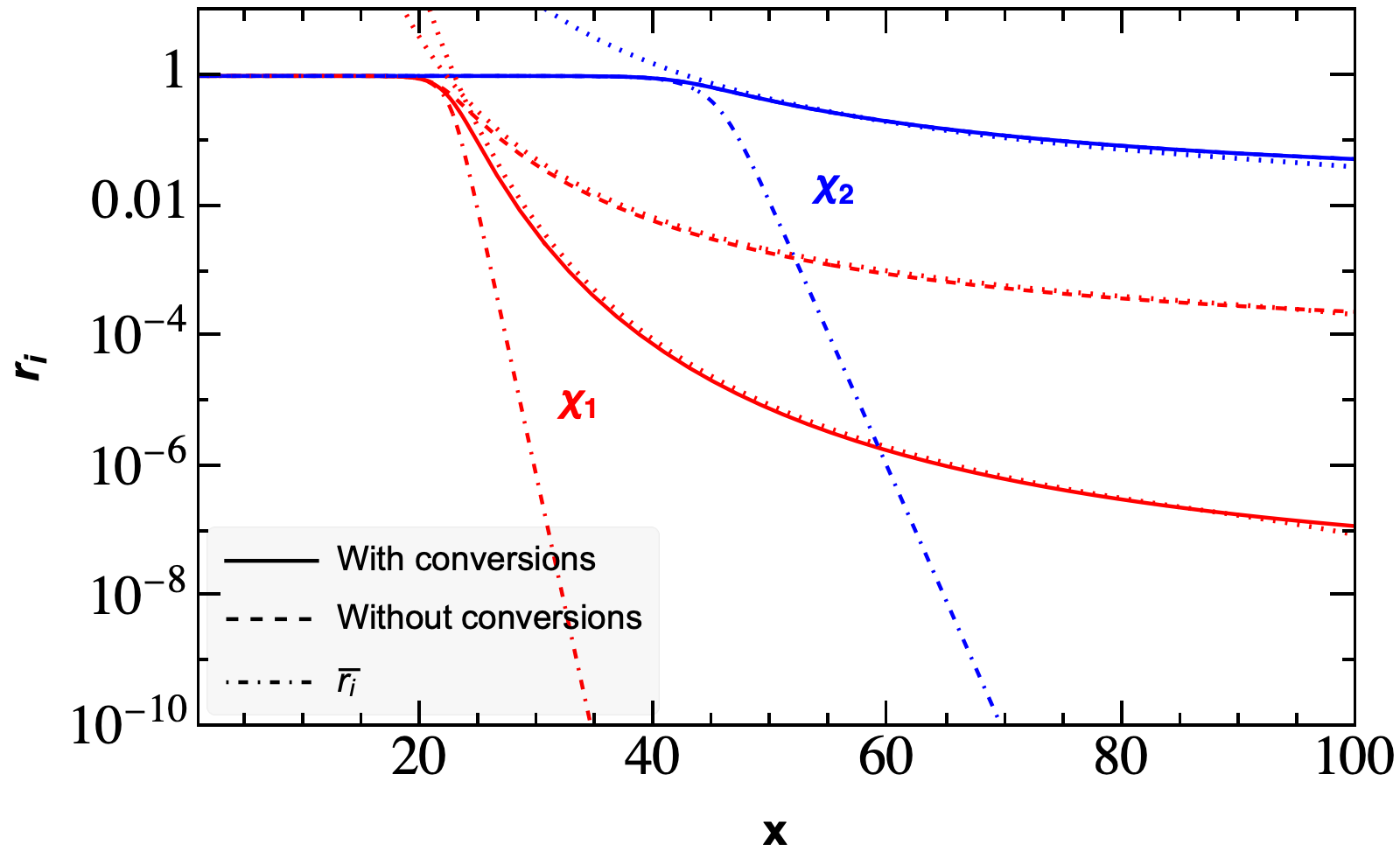}
\includegraphics[width=0.49\textwidth]{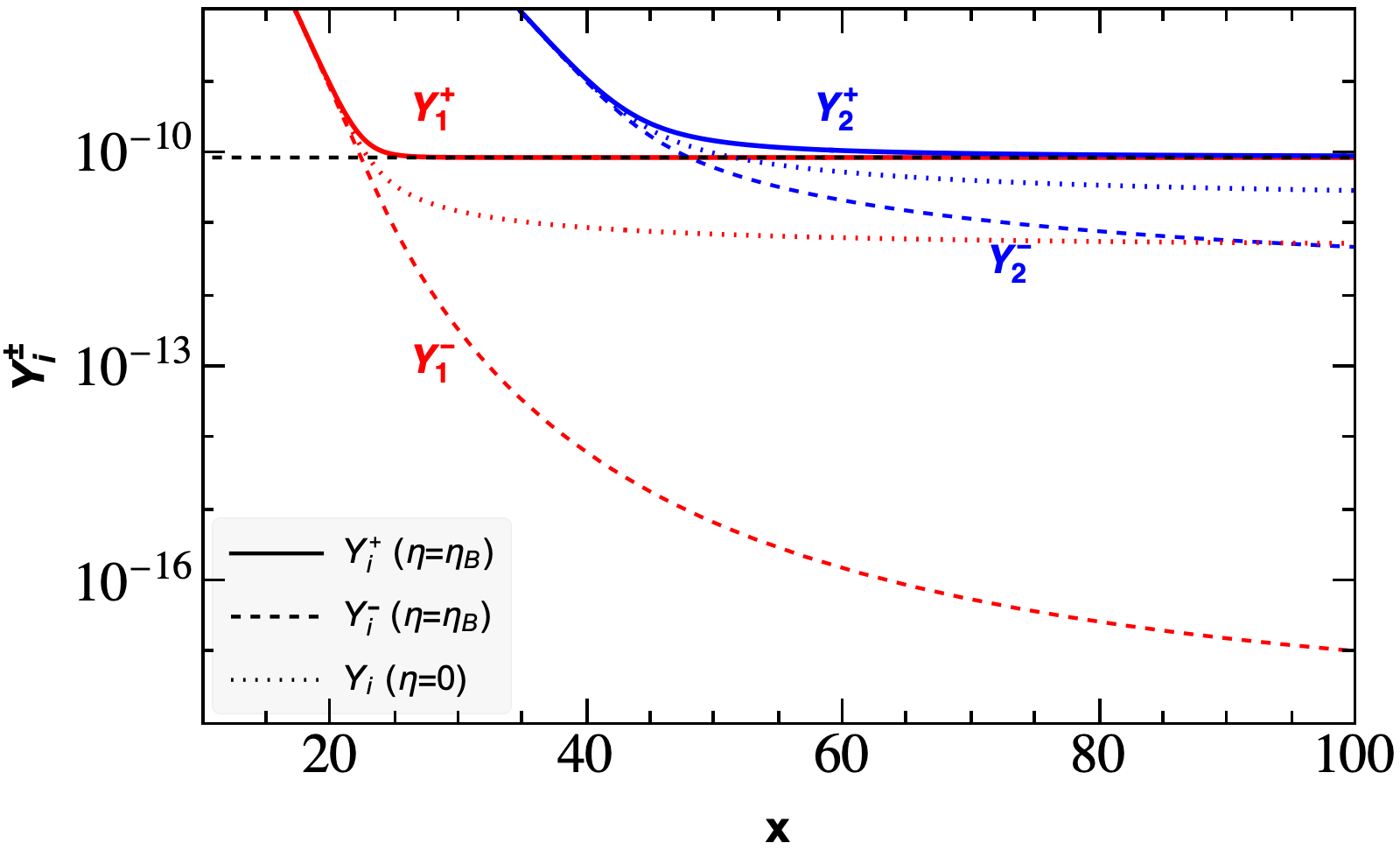}
\caption{Evolution of asymmetric ratios and yields for $\chi_1$ (red) and $\chi_2$ (blue) in the asymmetric case. We choose $m_1=20 {\rm~GeV},\,m_2=10 {\rm~GeV},\,\annone =\anntwo =\con = 2 {\rm~pb}$ and $\eta_{1,2}=\eta_B \sim 0.88 \times 10^{-10}$. \emph{Left panel:} Numerical solution for asymmetric ratios $r_i$ in presence (solid) and absence (dashed) of conversions. The dot-dashed curves represent the equilibrium distributions $\bar{r_i}(x)$. The dotted red and blue curves correspond to the analytic solution and coincides well with the numerical solution. \emph{Right panel:} The solid (dashed) curves correspond to $Y_i^+$ $(Y_i^-)$, and the dotted curves correspond to the symmetric case i.e. $\eta_{1,2}=0$. The black dashed line corresponds to the value of $\eta_B$.}
\label{fig:r}
\end{figure}

In Fig.~\ref{fig:r}, we plot the evolution of the asymmetric ratios and components after solving Eq.~\eqref{eq:B2} and using Eq.~\eqref{eq:ypym}.
Note that $Y_1^+ \sim Y_2^+ \sim \eta_B$, due to our choice of $\eta_1=\eta_2 =\eta_B$ and large annihilation cross-sections, so that the symmetric component is significantly depleted. In general, one obtains $Y_i^+ \sim \eta_i$. From the left plot in Fig.~\ref{fig:r}, it can be seen that the heavier component, $\chi_1$, is more asymmetric than the lighter one, $\chi_2$,  without conversions. Including conversions significantly reduces the symmetric component of the heaviest state. This can be seen in the right plot where $Y_i^{+}-Y_i^{-} \sim \eta_B~(i=1,2)$ but $Y_{1}^{-}\ll Y_2^{-}$. Hence, we observe that while the heavier component is mostly asymmetric, the lighter one is mostly symmetric. Notice that the analytic solutions derived in Sec. \ref{sec:an} are contrasted with the numerical solutions in Fig.~\ref{fig:r} (dotted lines) and are found to be in agreement.

\begin{figure}[!htb]
\centering
\includegraphics[width=.49\linewidth]{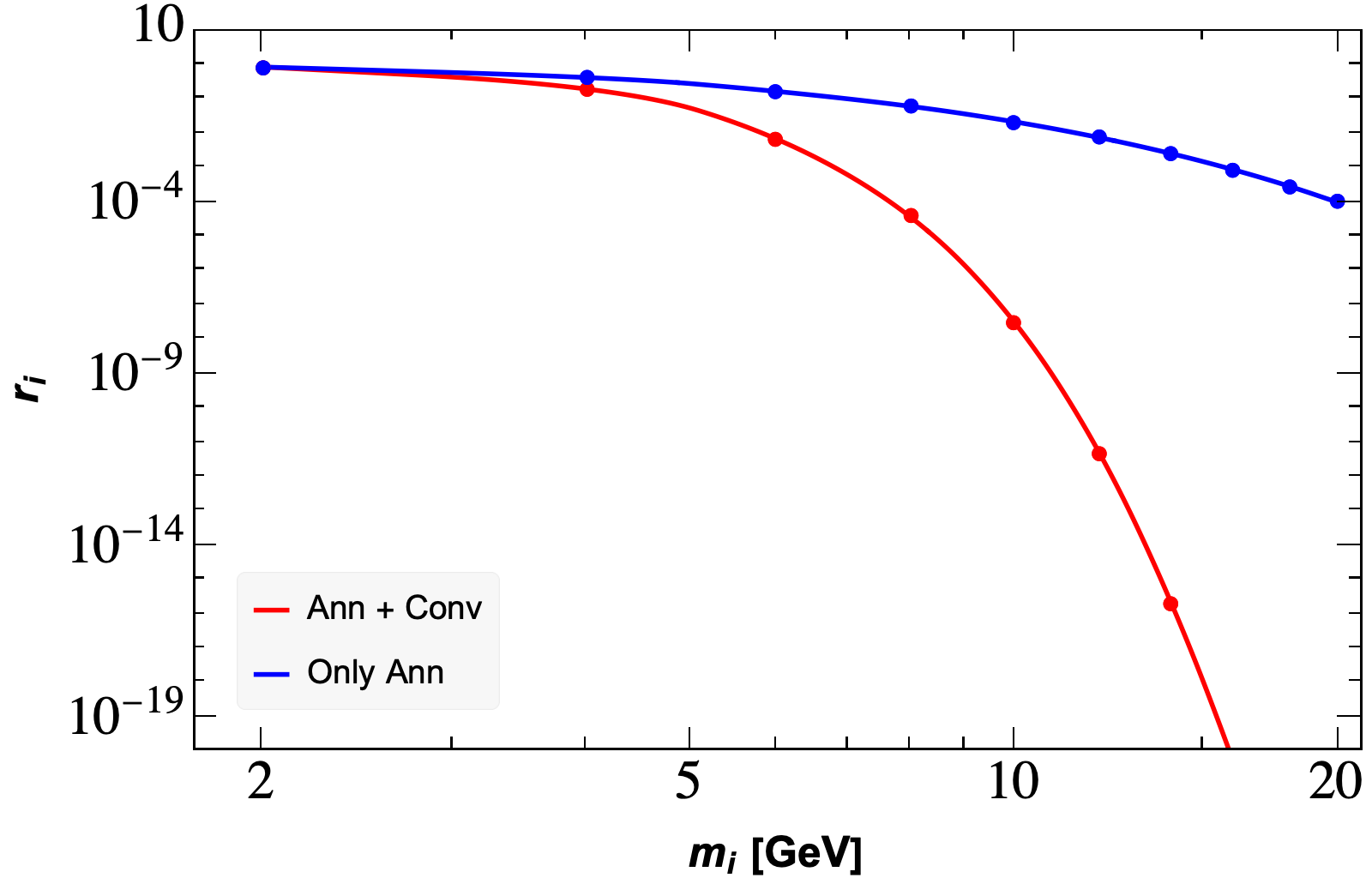}
\includegraphics[width=.49\linewidth]{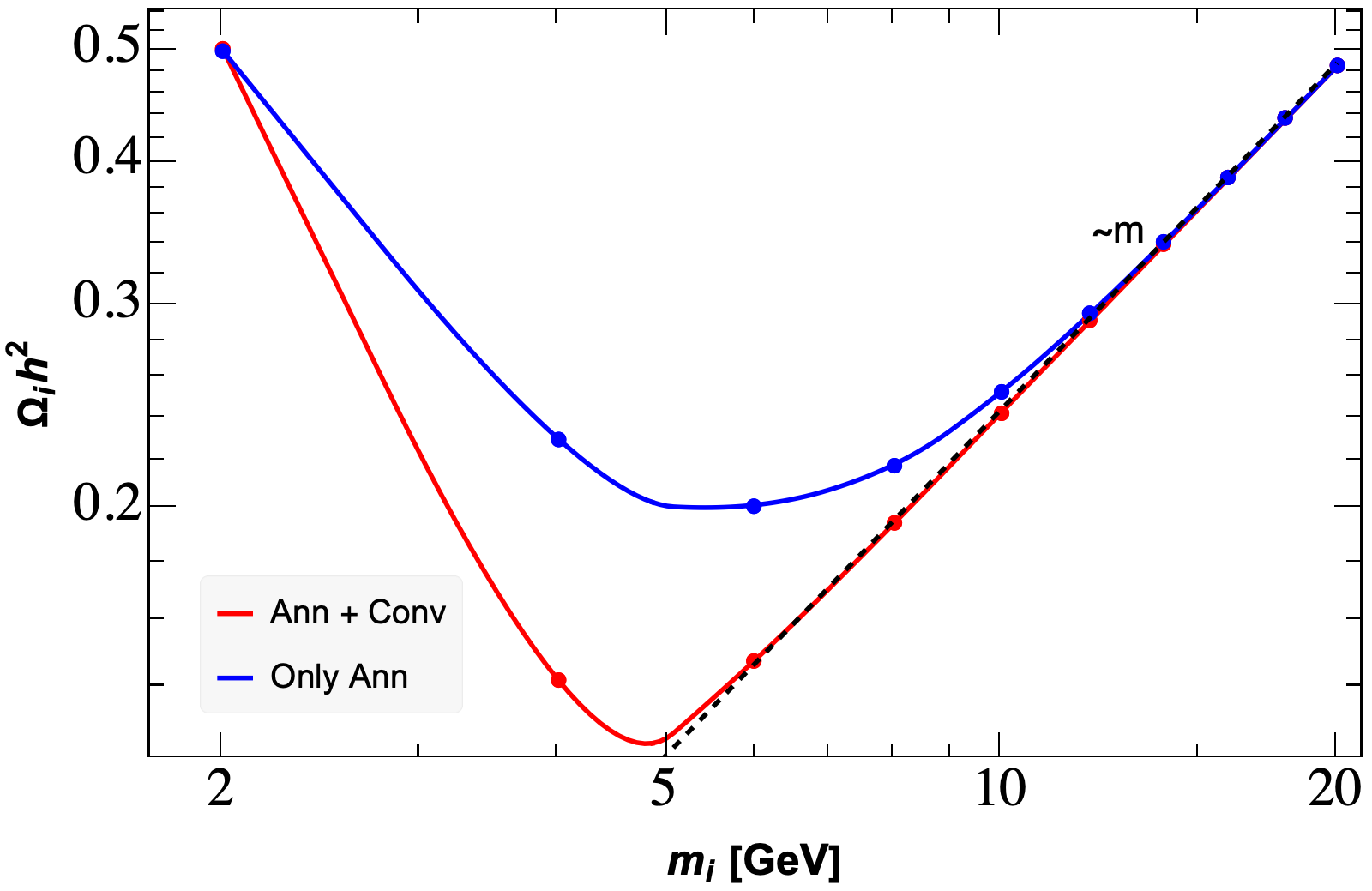}
\caption{\emph{Left panel:} Evolution of asymmetric ratios for 10 DM particles with $m_i - m_{i+1} = 2 {\rm~GeV}$, $m_1= 20{\rm~GeV}$, $\braket{\sigma v} =2{\rm~pb}$ and $\eta_{i} = \eta_B$, in presence (red) and absence (blue) of conversions. \emph{Right panel:} The individual abundance of DM components versus mass in the asymmetric case. The black dashed curve depicts the direct proportionality to mass as particles become more asymmetric i.e. $r_i \r 0$.}
\label{fig:r10}
\end{figure}

Now we extend the analysis to $N=10$ components to show the effects of conversions and annihilations, in a similar way to the analysis of symmetric components done above, keeping the value of annihilation and conversion cross sections equal for all components and fixing the initial asymmetries equal to the baryon asymmetry $\eta_B$. In Fig.~\ref{fig:r10}, we show the individual asymmetric ratios (left panel) and abundances (right panel).

Here, unlike the symmetric case, one cannot fit the asymmetry ratios as power laws in the DM mass. For the case without conversions, we find $r_i \sim a\, e^{-b\, m_i}$ is a good fit whereas when conversions are included, we can fit the individual asymmetric ratios as the product of a power law and an exponential: $r_i \sim a'\, m_i^{b'} e^{-c\, m_i}$, where $a',b',c$ depend on the value of cross sections. Notice that the DM mass enters subtly via $x_{f_i} = m_i/T_i$ in the expression for $\Phi$. We find that for similar values of the cross sections, the asymmetry ratios for the lighter species are larger than that of the heavier ones. This implies that, for a given initial asymmetry $\eta$, the heavier species are more asymmetric than the lighter ones, and they become even more asymmetric in the presence of conversions, thus they can be considered to be fully asymmetric.

From the right plot of Fig.~\ref{fig:r10}, it can be seen that the individual abundances coincide for the heavier particles with and without conversions and scale with the DM mass $m_i$. Since the heavier species can be considered to be fully asymmetric, their abundances are independent of conversion and annihilation cross sections. This is valid if the cross-sections are larger than a certain value. For example, in this case, annihilation cross-sections larger than $\mathcal{O}(0.1){\rm~pb}$ lead to $r < 0.1$.

As we can see, the presence of conversions reduces the total abundance also in the partially-asymmetric case. For initial asymmetries close to the baryonic one, i.e., $\eta_i \simeq \eta_B$, we find that only in the case of purely asymmetric DM states with masses above $\sim 10$ GeV, is the effect of conversions irrelevant. Furthermore, the direct proportionality of abundance with mass is a good approximation for even lower masses in the presence of conversions, which tend to make the species more asymmetric, as illustrated by the black dashed curve. We also see that the abundance is at or below current Planck values for DM masses around 5 GeV, as expected in asymmetric DM models (see Section~\ref{sec:asymmetry}) due to the initial asymmetry being equal to the baryonic one.

\subsection{Total relic abundance \label{sec:relic}}

We have seen that it is natural that if, DM has different states, the individual symmetric and asymmetric components can contribute significantly to the relic abundance for reasonable values of the cross sections. We follow closely the notation in Ref.~\cite{Graesser:2011wi}.

For the case with initial asymmetries, the total DM abundance will consist of both symmetric and asymmetric components and the DM is partially asymmetric, with an energy density given by
 \begin{equation}\label{eq:relic}
\rho_{\rm DM}  \equiv \sum_i \rho_{i} = s \sum_i m_i \,\eta_i \left(1+2\frac{r_{\infty,i}}{1-r_{\infty,i}}\right)\,,
 \end{equation}
where $r_{\infty,i}$ depends on $\Phi_i(\infty,m_i)$. The first term corresponds to the asymmetric components (present for $r_{\infty,i} \rightarrow 0$), and the second one to the symmetric ones (which scale as the inverse power of the sum of annihilation and conversion cross sections for $r_{\infty,i} \rightarrow 1$). 
Reproducing the relic abundance yields a constraint between the DM masses, the asymmetries and the annihilation and conversion cross sections through this relation. Notice that in some cases the initial asymmetries may be similar, $\eta_i \simeq \eta$, i.e., if they come from the same operator, or if conversions/annihilation are large, and we can factorize $\eta$ from the first term.

\begin{figure}[!htb]
\centering
\includegraphics[scale=0.4]{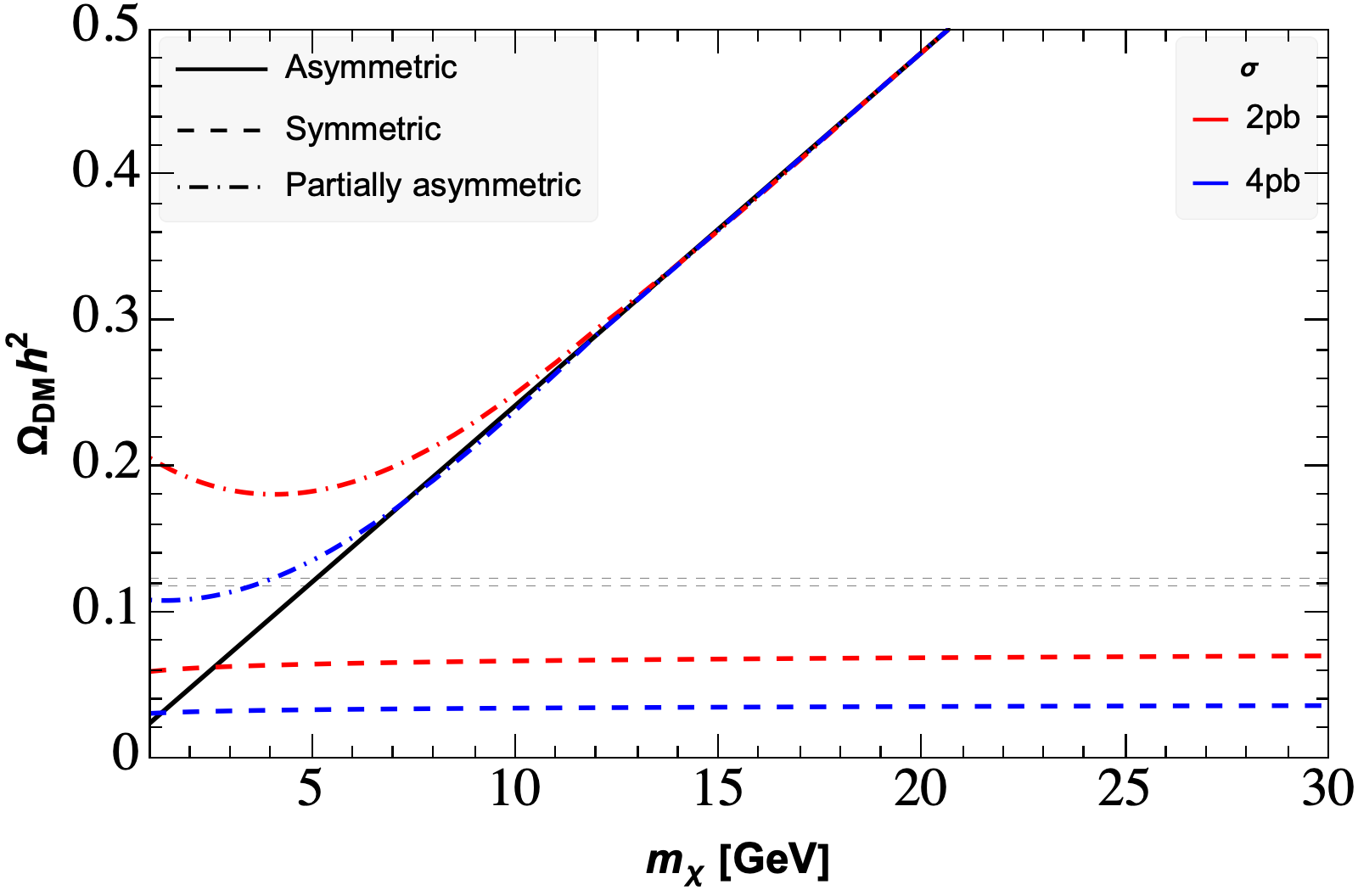}
\caption{Relic abundance as the function of DM mass for asymmetric (solid), symmetric (dashed) and partially-asymmetric (dot-dashed) cases. The red and blue curves correspond to two different choices of $\sigma_i =2 {\rm~pb},\, 4 {\rm~pb}$ respectively. The asymmetric case is linearly dependent on mass and $\eta$, which is chosen to be equal to $\eta_B$. The symmetric case is almost independent of mass and depends only on the values of $\sigma_i$. The partially-asymmetric case becomes independent of $\sigma_i$ for large mass values, e.g., it starts exhibiting a completely asymmetric behaviour.}
\label{fig:relcomp}
\end{figure}  

In the completely asymmetric case, i.e. $r_{\infty,i}=0$, the total DM abundance just depends on the masses and initial asymmetries,
\begin{equation}\label{eq:reladm}
\rho_{\rm ADM} = s\,\sum_i \eta_i\,m_i \,.
\end{equation}

In the symmetric limit $r_{\infty,i} \rightarrow 1$, the expansion of the second term gives
 \begin{equation}\label{eq:relsym}
\rho_{\rm SDM} \simeq s\,\sqrt{\frac{180}{\pi M^2_{\rm Pl}}} \sum_i \frac{1}{\, \sigma_i\,\Phi_i(\infty,m_i)}\,,
 \end{equation}
which is independent of the initial asymmetries and very mildly dependent on the DM mass via $\Phi_i(\infty,m_i)$. In presence of conversions, $\sigma_i$ has to be summed over the conversion cross sections as well, such that total cross-section becomes
\begin{equation}\label{eq:modsig}
\sigma_{i,t} = \langle \sigma_{\rm ann,\,i}\, v\rangle + \sum_{j>i} \braket{\sigma_{\rm conv,\,ij}\,v}\,.
\end{equation}

Using Eqs.~\eqref{eq:cosco} and \eqref{eq:relsym}, one can also re-express the fractional asymmetry for heavier components in presence of conversions as~\cite{Graesser:2011wi}
\begin{equation}
r_{i\,,\infty}=e^{-\lambda_{i,t}\, \eta_i \, \Phi_i}={\rm exp}\left[-2\left(\frac{\sigma_{i,t}}{\sigma_{\rm WIMP}}\right)\left(\frac{\Phi_{i,t}}{\Phi_{\rm WIMP}}\right)\frac{1-r_{i\,,\infty}}{1+r_{i\,,\infty}}\right]\,,
\end{equation}
where $\lambda_{i,t} =(\pi/45)^{1/2}M_{\rm pl}m_1 (\sigma_i+\sigma_c)$, and $\sigma_{\rm WIMP}$ and $\Phi_{\rm WIMP}$ correspond to a symmetric DM particle of the same mass, $m_i$. It should be noted that the freeze-out temperature $(x_{f_i})$ entering the expression for $\Phi$ in Eq.~\eqref{eq:phi} is modified when conversions are included, hence we use $\Phi_{i,t}$.

Naively, the relic abundance in the partially-asymmetric case (Eq.~\eqref{eq:relic}) might be expected to be the sum of Eqs.~\eqref{eq:reladm} and \eqref{eq:relsym}; however, that's not the case as the symmetric component also depends on $\eta$ \cite{Graesser:2011wi}. Only for scenarios where the DM components are completely symmetric or asymmetric, the total abundance can be given by their sum
\begin{equation}
\frac{\rho_{\rm DM}}{s} \simeq  \sum_{i={\rm ADM}} \eta_i\, m_i +\left(\frac{180}{\pi M^2_{\rm Pl}}\right)^{1/2} \sum_{j={\rm SDM}} \frac{1}{\sigma_j\,\Phi(\infty,m_j)}\, \,,
\end{equation}
for $i \neq j$. 
 
In Fig.~\ref{fig:relcomp}, we plot the dependence of the relic abundance on masses\footnote{The plot is valid for 1DM case or for multi-component DM without conversions.} and cross sections  to show the differences between the symmetric, asymmetric and partially-asymmetric cases. It can be seen that the curve corresponding to partially asymmetric is not equivalent to the sum of the other two, and starts tracking the asymmetric curve for larger masses. The mass at which it starts tracking the asymmetric case depends on the value of $\sigma_i$ and $\eta_i$.

\section{Effective Parameterisation \label{sec:numerics}}

In the analysis for symmetric and asymmetric components performed above, we have considered all cross sections to be independent of DM mass. However, these are model dependent and in particular, also scale with the masses of the particles involved. For instance, considering only \textit{s}-wave annihilations we can parameterise the DM annihilation cross section at freeze-out as 
\begin{equation}
\braket{\sigma_{\rm ann,i}\,v} \simeq \l_i^2\, \frac{m_i^2}{(m_i^2 + \Lambda^2)^2}\,,
\end{equation}
where $m_i$ is the DM mass, $\lambda_i$ the coupling characterising the DM-SM interaction and $\Lambda$ is the mass scale of the heavy mediator involved. We consider two limiting cases for our analysis,
\begin{align}\label{eq:miannv}
\braket{\sigma_{\rm ann,\,i}\,v} \sim
\begin{cases}
 {a_i}/{m_i^2} {\rm~for~} \Lambda \ll m_i \\
{b_i\,m_i^2}/{\Lambda^4} {\rm~for~} \Lambda \gg m_i
\end{cases}\,,
\end{align}
corresponding to light and heavy mediator respectively, with $a_i$ and $b_i$ being model-dependent numerical parameters. In a similar manner, we parameterise the conversion cross section as
\begin{equation}
\braket{\sigma_{\rm conv,\,ij}\,v} \simeq \l_{ij}^2\, \frac{m_i^2}{(m_i^2 + \Lambda^2)^2}\,\sqrt{1-\frac{m_j^2}{m_i^2}}\,,
\end{equation} 
where $\l_{ij}$ is the interaction strength between the DM components with masses $m_{i}$ and $m_j$. As before, we consider the limiting cases
\begin{align}\label{eq:miconv}
\braket{\sigma_{\rm conv,\,ij}\,v} \sim
\begin{cases}
c_{ij}\,\frac{ (m_i^2 -m_j^2)^{1/2}}{m_i^3} {\rm~for~} \Lambda \ll m_i \\
d_{ij}\,\frac{ m_i (m_i^2 -m_j^2)^{1/2}}{\Lambda^4} {\rm~for~} \Lambda \gg m_i\,.
\end{cases}
\end{align}

Hence, in order to describe the dynamics of $N$-component DM, we require $N$ DM masses, $N$ DM-SM couplings and $N(N-1)/2$ DM-DM couplings i.e. $N(N+3)/2$ parameters in total. In the following, we analyse the results in a model independent way, treating $a,b,c,d$ as free parameters and fixing them to be equal for all $i,j$. We then provide an example model in Appendix~\ref{sec:model}.

\subsection{Symmetric components \label{sec:symnum}}

\subsubsection{Light mediators \label{sec:lightmed}}

First, let us consider the case of light mediators, where the cross-sections are independent of the mediator mass, see first line of Eqs.~\eqref{eq:miannv} and \eqref{eq:miconv}, and depend on the new parameters $a_i$ and $c_{ij}$, taking $N =10$ and varying the DM masses from $300{\rm~GeV}$ to $1200{\rm~GeV}$. In the absence of conversions ($c_{ij}=0$), all Boltzmann Equations are decoupled and we roughly obtain the individual yields (abundances) scaling as $\sim m_i$ ($\sim m_i^2$). This is expected since $Y_i$ scales as $\sim m^{-1}$ when we treat all cross sections to be similar and independent of mass. 

\begin{figure}[!htb]
\centering
\includegraphics[width=.49\linewidth]{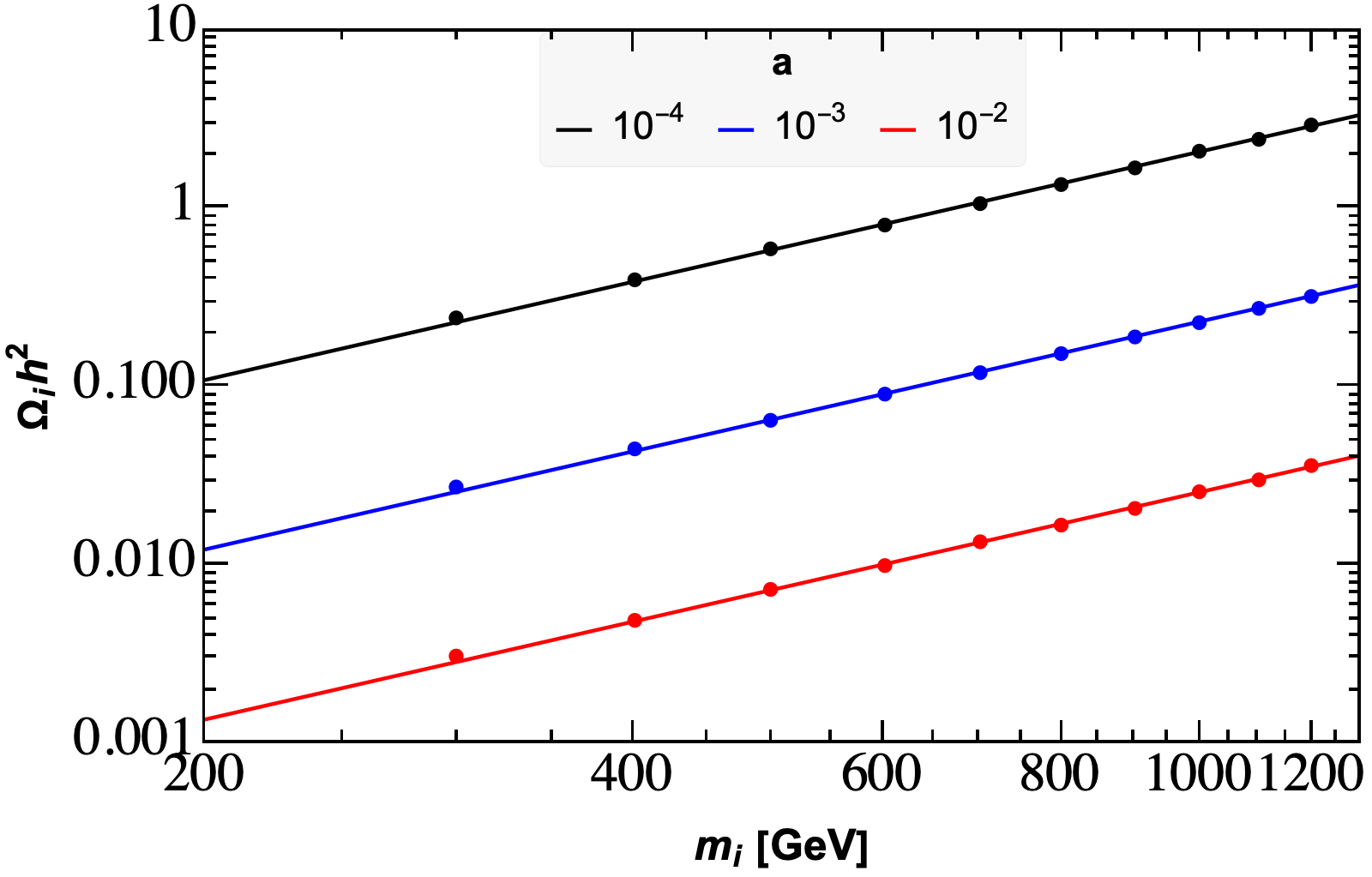}
\includegraphics[width=.49\linewidth]{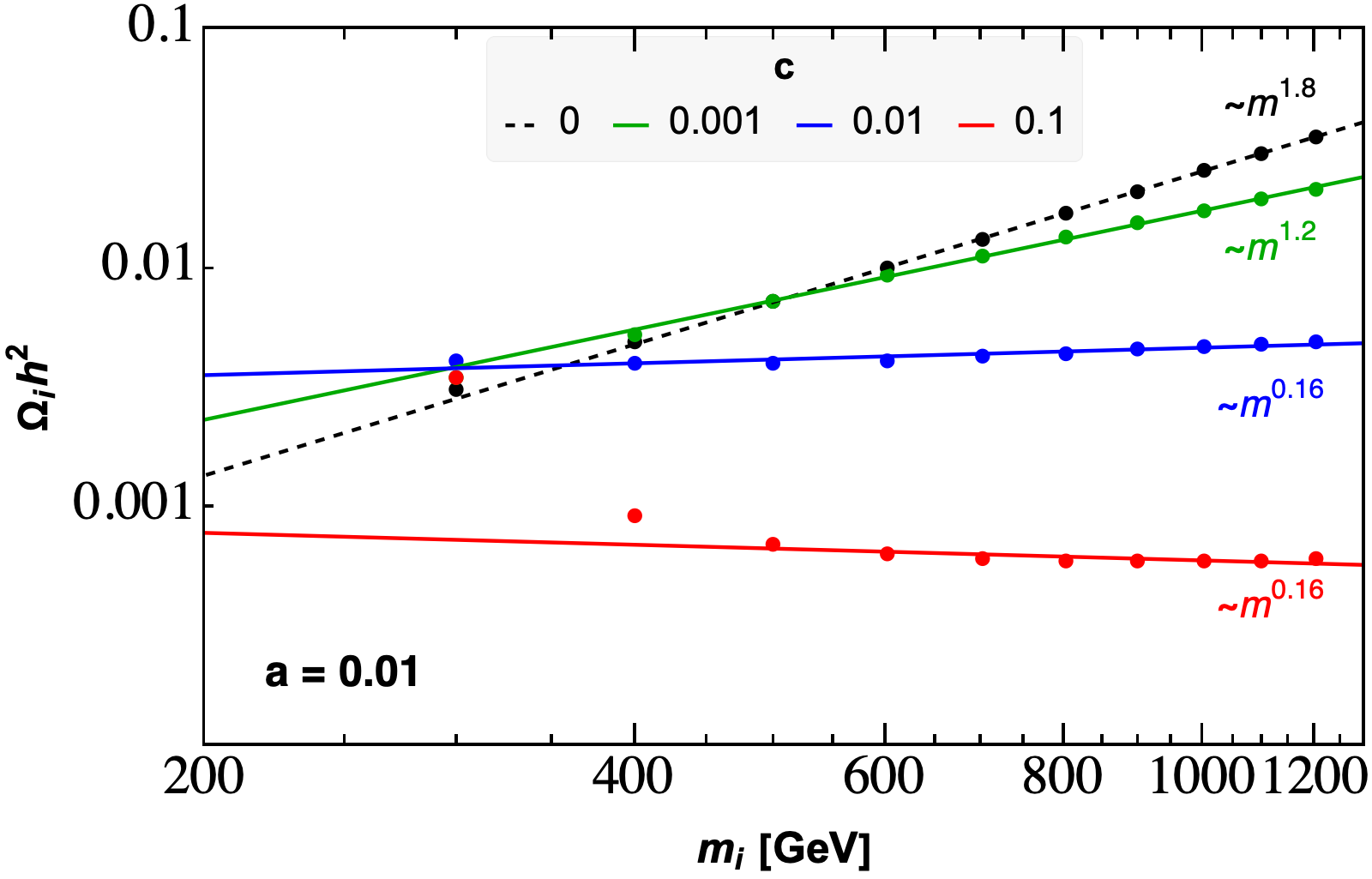}
\caption{Individual abundances with DM mass in \emph{the light mediator case,} without (\emph{left panel}) and with (\emph{right panel}) conversions, for $N=10$.}
\label{fig:lmann}
\end{figure} 

The asymptotic abundances are plotted as a function of mass in the left plot of Fig.~\ref{fig:lmann}, for different choices of $a$. In this case, the abundances follow a power-law of the form
\begin{align}
\Omega_i h^2 \sim \frac{\kappa}{a}\, m_i^{1.8}\,,
\end{align}
where $\kappa$ is to be determined from the fits.

Moreover, we find that the scenario becomes interesting in presence of conversions with an additional dependence on the parameter $c$ (we take all $c_{ij} \equiv c$).  The total conversion cross sections of a given DM particle into the lighter states $\braket{\sigma_{\rm conv,\,i}\,v} = \sum_{j>i} \braket{\sigma_{\rm conv,\,ij}\,v}$ are of similar size. This is due to the fact that while heavier DM particles have more DM states to convert into, their conversion cross sections are suppressed by their mass, whereas lighter states have less DM states to convert into, but their cross sections are much larger. These effects partially compensate each other. As a result of this, the power law dependence on mass for abundances varies depending on the strength of conversions with respect to annihilations, unlike the case of mass-independent cross sections, where we saw $\Omega_i h^2 \propto m_i^{-1}$ in presence of conversions. We illustrate this in the right plot of Fig.~\ref{fig:lmann}. As can be seen, for $c \gtrsim a$, the abundances do not change much with the mass of the DM component. The abundances in this case can be expressed in the form
\begin{align}\label{eq:lmy}
\Omega_i h^2 \sim \frac{\kappa^\prime}{\left(\frac{a}{m_i^2}+ \delta c \right)}\,,
\end{align}
where $\kappa^\prime$ and $\delta$ are numbers determined from the fit. If the conversions are weak, i.e. $c \ll a$, the first term in the denominator of Eq.~\eqref{eq:lmy} dominates and $\Omega_i h^2 \propto m_i^2$. On the other hand, for significant conversions i.e. $c \sim a$ or $c \gg a$, the second term in the denominator dominates and we find $\Omega_i h^2$ does not depend on $m_i$. Therefore with conversions, $\Omega_i h^2 \sim\,\kappa^{\prime}\, m^{\beta}$ with $0 \lesssim \beta \lesssim 2$. 

Note that for extremely large conversions, the fits are determined by ignoring the lightest and next-to-lightest DM particles as they do not follow the same mass-suppression as heavier ones. However, for moderate to weak conversions, we find that it is possible to perform a fit for all the DM components. 

\subsubsection{Heavy mediators \label{sec:heavymed}}

In the case of heavy mediators and/or light DM, see second line of Eqs.~\eqref{eq:miannv} and \eqref{eq:miconv}, where annihilations are proportional to $b_i$ and conversions to $d_{ij}$ (we take all equal), the cross sections are proportional to the squares of DM masses and the individual yields and abundances are thus expected to scale as $\sim m_i^{-3}$ and $\sim m_i^{-2}$, respectively. 

\begin{figure}[!htb]
\centering
\includegraphics[width=.49\linewidth]{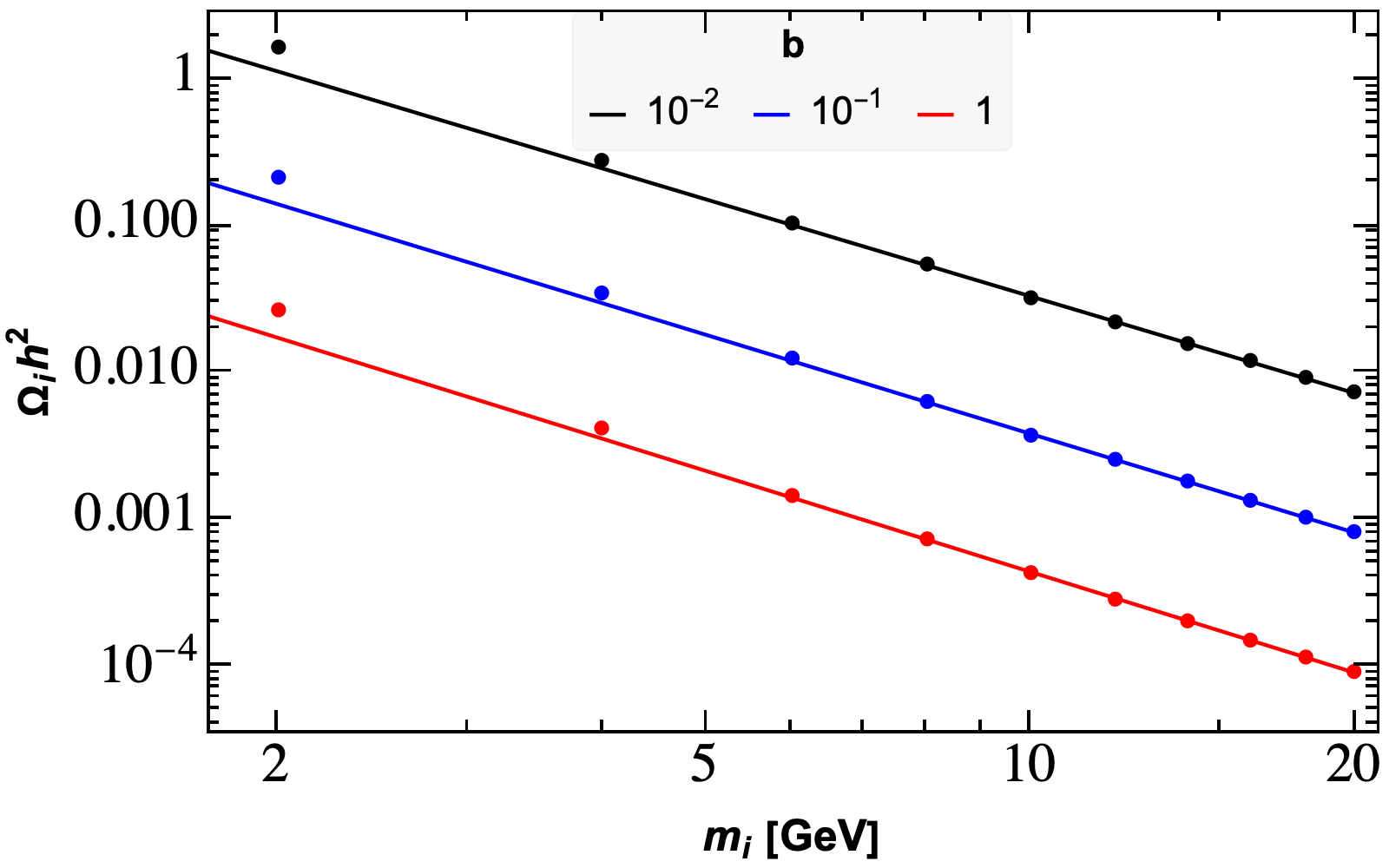}
\includegraphics[width=.49\linewidth]{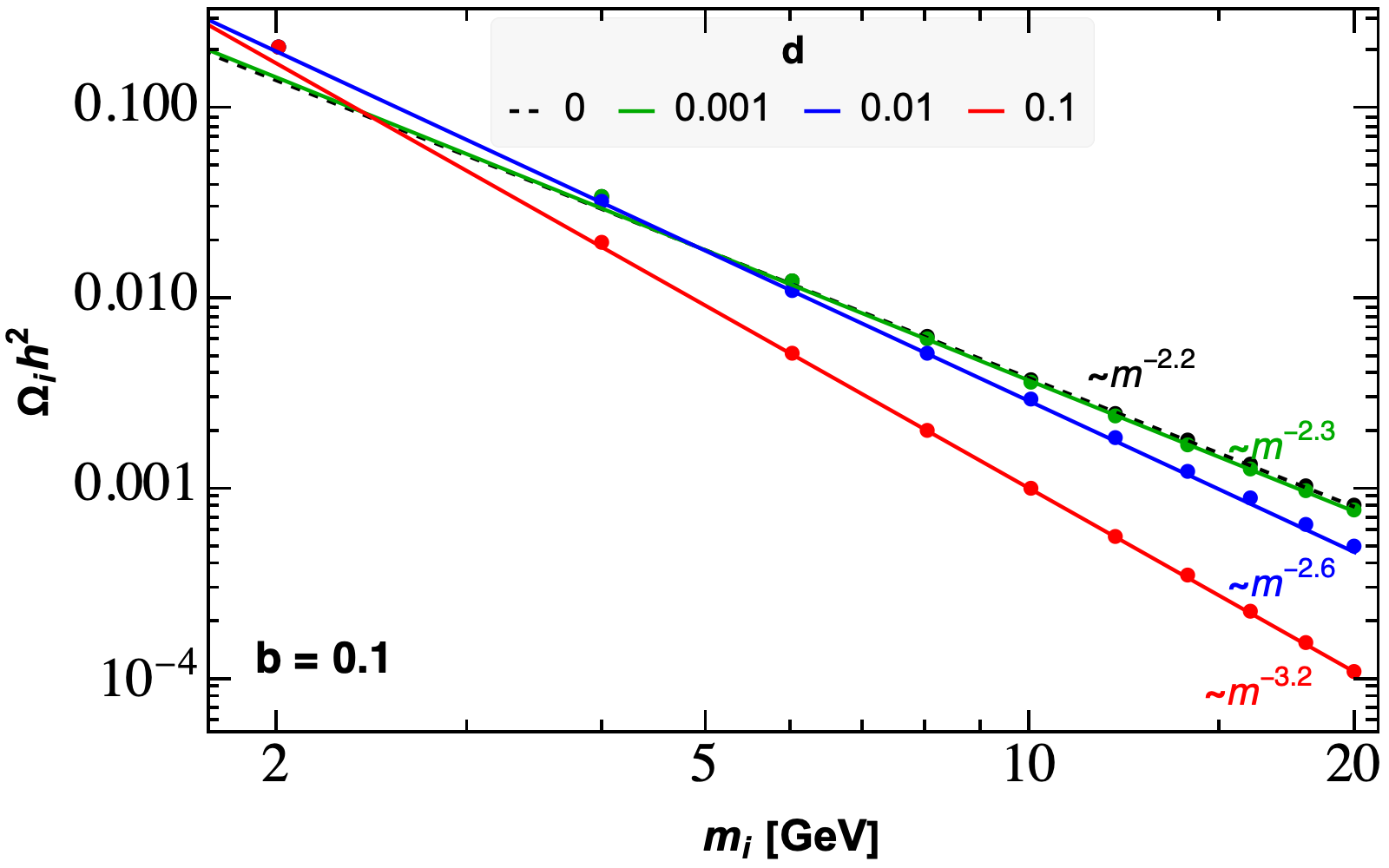}
\caption{Individual abundances with DM mass in \emph{the heavy mediator case}, without (\emph{left panel}) and with (\emph{right panel}) conversions, for $N =10$, with $\Lambda \sim 100{\rm~GeV}$.}
\label{fig:hmann}
\end{figure} 

The case of no conversions is shown in the left plot of Fig.~\ref{fig:hmann} for different choices of the parameter $b$. We have taken the heaviest DM component mass to be $20{\rm~GeV}$ and the lightest $2 {\rm~GeV}$. We also fix the mediator scale to $\Lambda =100 {\rm~GeV}$. The asymptotic relic abundances follow a power law of the following form
 \begin{align}
\Omega_i \sim \frac{\kappa}{b}\, m_i^{-2.2}\,,
\end{align} 
where $\kappa$ is determined from the numerical fits.

In presence of conversions, the dependence on mass will vary depending on the strength of conversions with respect to annihilations, similar to the light mediators case. However, due to the direct proportionality to mass, the total conversion cross sections of each species are not similar unlike in the previous case, and scale as $\sim m_i^3$. Therefore, we can express the abundances as
\begin{align}\label{eq:hmy}
\Omega_i h^2 \sim \frac{\kappa^\prime \Lambda^4}{ \left({b}\,{m_i^2}+ \delta\, d\, m_i^3 \right)}\,,
\end{align}
where $\kappa^\prime, \delta$ are determined from the numerical fits. When conversions dominate, i.e. $d \gtrsim b$, we get $\Omega_i h^2 \propto m_i^{-3}$, whereas for weak conversions, we obtain roughly $\Omega_i h^2 \propto m_i^{-2}$. Hence, for the heavy mediators case, we have $\Omega_i h^2 \sim \kappa^\prime\, m^{\gamma}$ with $-3 \lesssim \gamma \lesssim -2$, see the right plot of Fig.~\ref{fig:hmann}.

\subsection{Asymmetric components \label{sec:asymnum}} 

As discussed above, in the case of completely asymmetric dark matter, the relic abundance is independent of the annihilation and conversion cross sections, whereas there is a slight dependence on these cross sections in the case of partially-asymmetric dark matter. 

From the analytic expressions derived above, we know that the asymmetry ratio is exponentially suppressed by the annihilation cross section and the initial asymmetry $\eta_i$. In the following, we discuss the effect this has on the DM relic abundance for the light and heavy mediator cases.

\subsubsection{Light mediators \label{sec:aslightmed}}

\begin{figure}[!htb]
\centering
\includegraphics[width=.49\linewidth]{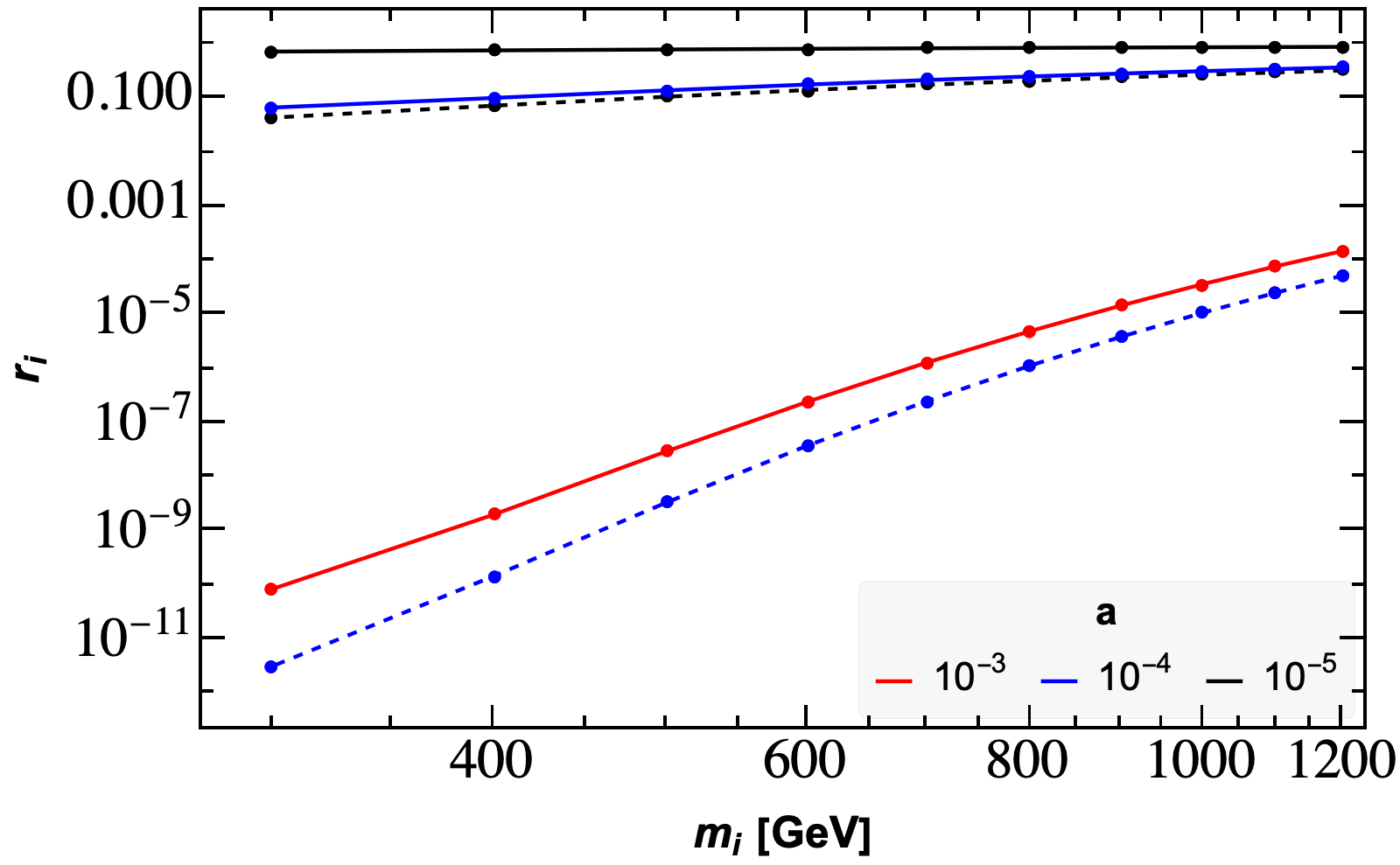}
\includegraphics[width=.49\linewidth]{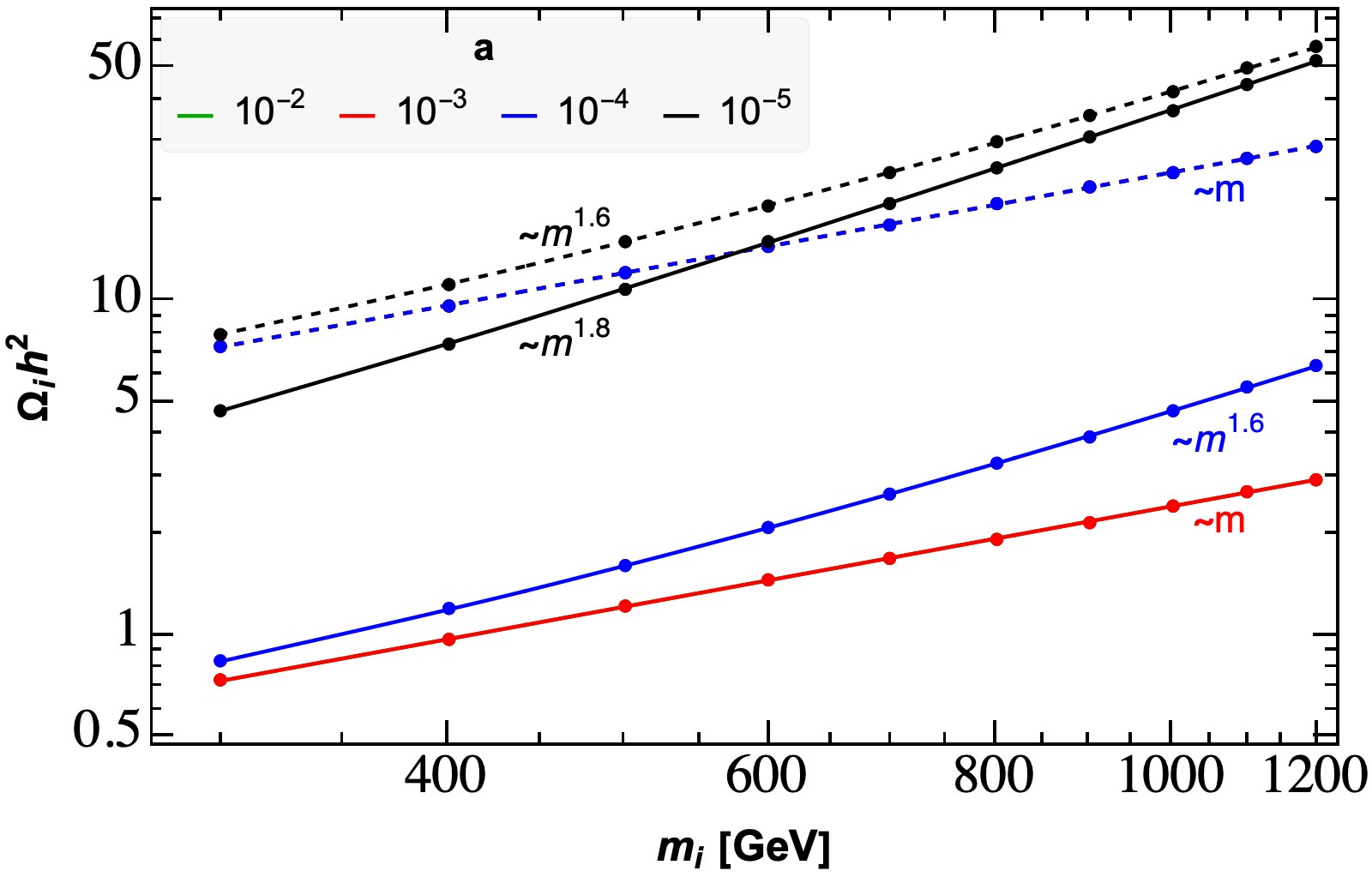}
\caption{Asymmetry ratio (\emph{left panel}) and abundance (\emph{right panel}) versus DM mass for different choices of $a$ in \emph{the light mediator case}, without conversions, for $N =10$. \emph{Left:} Notice that all the dashed curves $(\epsilon = 1)$ are suppressed compared to the solid ($\epsilon =0.1$) ones. The red dashed curve is highly suppressed and does not appear in the plot. $\emph{Right:}$ The green solid curve corresponding to $a=10^{-2}$ coincides exactly with the red solid curve, whereas the dashed green and red curves coincide exactly with the blue dashed curve.}
\label{fig:admlm}
\end{figure}

Let us first consider the case without conversions. When the cross sections are inversely proportional to the square of the DM masses, the lighter species will have a larger cross section and their asymmetry ratio gets exponentially suppressed compared to the heavier ones. The trends vary depending on the value of $\epsilon$ in Eq.~\eqref{eq:eps} and $a$ in Eq.~\eqref{eq:miannv}, which enter in Eq.~\eqref{eq:analytic}. In Fig.~\ref{fig:admlm}, we show the fractional asymmetries versus mass for different choices of $\eta$ (e.g. $\epsilon$, see Eq.~\eqref{eq:eps}). 

In the left plot of Fig.~\ref{fig:admlm}, it can be seen that the solid curves ($\epsilon = 0.1$) are larger than the dashed ($\epsilon=1$) ones for the same value of $a$, due to the suppression by $\eta$. At the same time, curves corresponding to larger values of $a$ have $r \ll 1$ and can be treated as completely asymmetric, while the curves corresponding to tiny values of $a$ are close to one, and are mostly symmetric. This is reflected in the relic abundance of the species, where we see that for larger values of $a$, the curves coincide and are a linear function of $m$ (i.e., the solid red and the dashed blue curves), whereas for smaller values of $a$, the abundances scale as $\sim m_i^{2}$, similar to what we observed for the symmetric components. 

\subsubsection{Heavy mediators \label{sec:aslightmed}}

\begin{figure}[!htb]
\centering
\includegraphics[width=.49\linewidth]{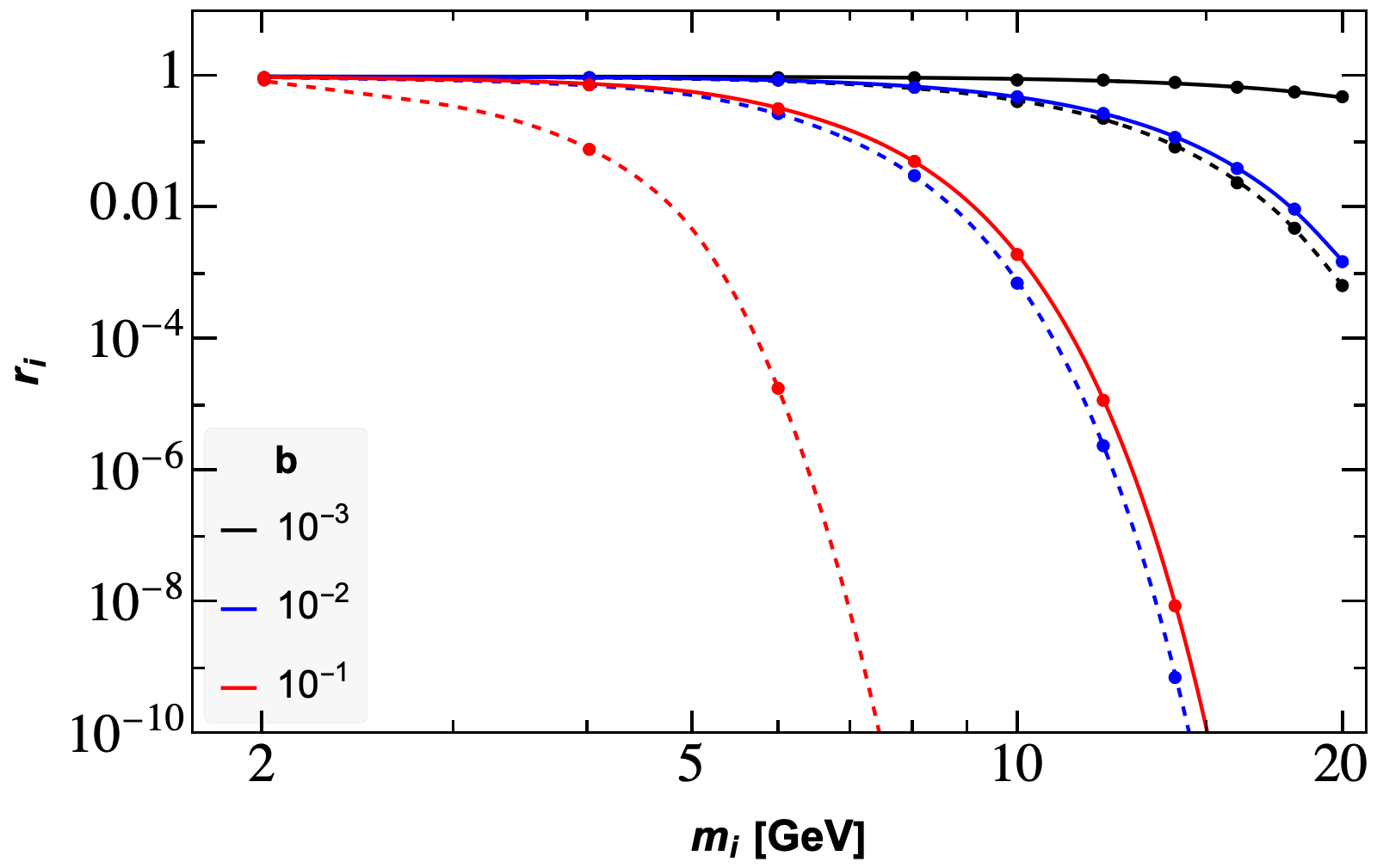}
\includegraphics[width=.49\linewidth]{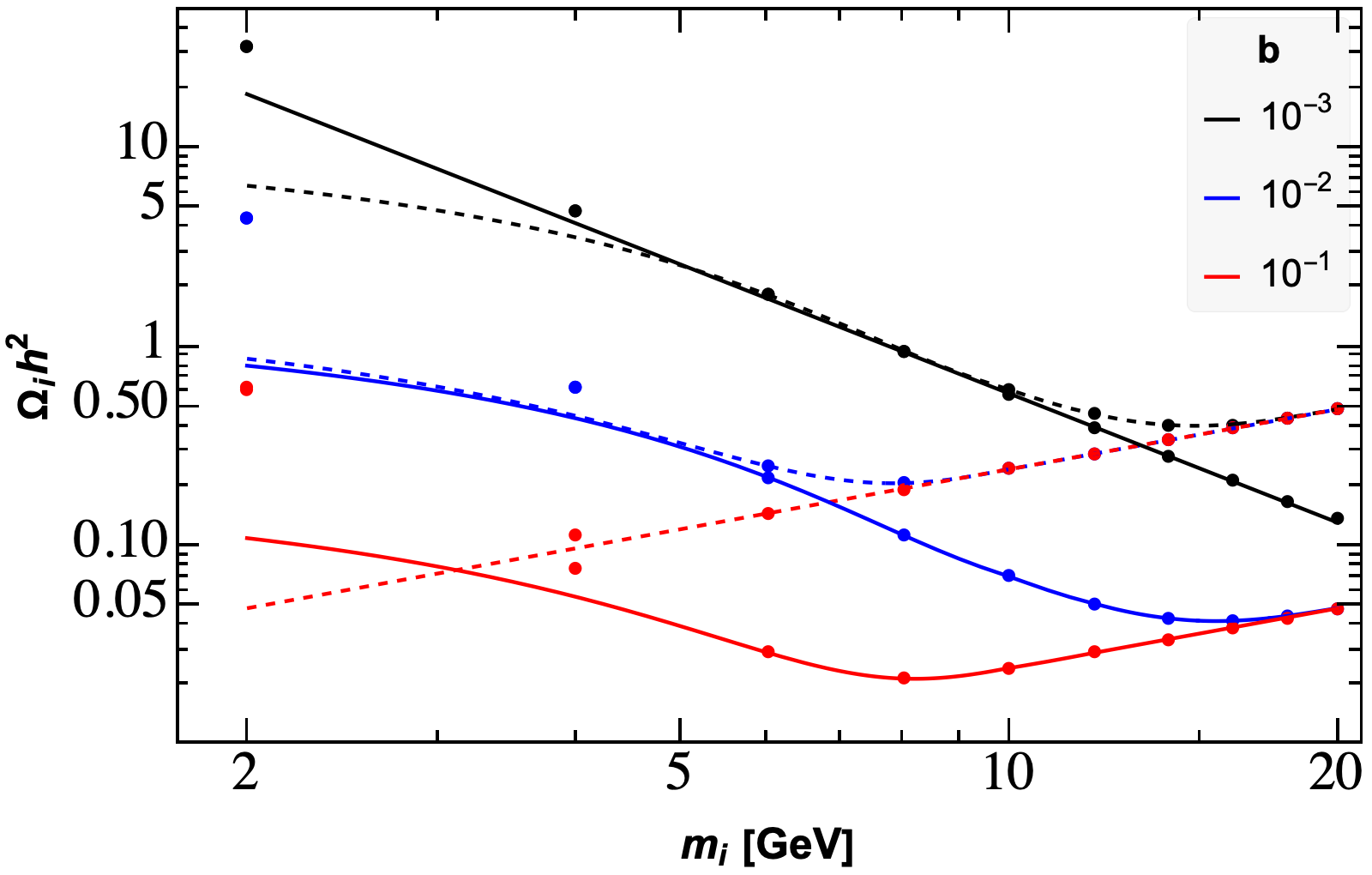}
\caption{Asymmetry ratio (\emph{left panel}) and abundance (\emph{right panel}) versus DM mass for different choices of $b$ in \emph{the heavy mediator case}, without conversions, for $N =10$, with $\epsilon = 1$ (dashed) and $\epsilon =0.1$ (solid).}
\label{fig:admhm}
\end{figure}

In the case of heavy mediators, as heavier components have larger cross sections, they are expected to be highly asymmetric in nature compared to the lighter ones. In Fig.~\ref{fig:admhm}, we show the asymmetry ratios and abundances for different choices of $b$ (annihilations) and $\epsilon$. We fix the mediator mass scale at $100 {\rm~GeV}$ and vary the DM masses in the range $2 - 20$ GeV. 
Comparing the asymmetry ratios with the light mediator case, c.f. Fig.~\ref{fig:admlm}, we see a reversed behaviour and now heavier particles in the set-up tend to be more asymmetric than the lighter ones. The suppression due to $\epsilon$ is evident in the left plot. Depending on the strength of annihilations, the abundance of heavier particles that are completely asymmetric is proportional to their mass whereas the abundance of the lighter ones roughly scales as $\sim m^{-2}$, similar to what we had seen in the symmetric case. Notice also how, the smaller the value of $\epsilon$, the larger the value of the DM mass where the abundance changes behaviour from mainly being symmetric-dominated to being asymmetric-dominated.

\subsubsection{Effect of conversions \label{sec:convadm}}

Here we discuss the effect of including conversions for both cases above. Since conversions deplete the symmetric components, the heavier particles which can convert to the lighter ones become more asymmetric, compared to the case of no conversions. However, the heavier states may be more symmetric (asymmetric) than the lighter ones in the case of light (heavy) mediators (c.f the left panels in Figs.~\ref{fig:admlm} and~\ref{fig:admhm}). Therefore, in presence of strong conversions, the abundance of the heavier species scales proportional to their mass and the initial asymmetry. For weak conversions, the behavior is similar to the case of only annihilations. Depending on the strength of conversions versus annihilations, one can categorise the multi-component asymmetric DM set-up as shown in Table~\ref{tab:convadm}. The only case where the DM can be treated as partially asymmetric is when both annihilations and conversions are weak; all other combinations lead to a completely asymmetric behavior. 

\begin{table}[!htb]
\centering
\begin{tabular}{|c||c|c|}
\hline
Annihilation/Conversion & \textit{Weak} & \textit{Strong}\\
 \hline \hline
 \textit{Weak} & Partially asymmetric & Asymmetric \\
 \hline
 \textit{Strong} & Asymmetric & Asymmetric \\
 \hline
\end{tabular}
\caption{\label{tab:convadm} Behavior of multi-component DM (except for the lightest component), depending on interaction strength. `Asymmetric' denotes that the DM is completely asymmetric.}
\end{table}

\subsection{Mass-dependence: power-law scaling \label{sec:powerlaw}}

For the parameterisations of annihilation and conversion cross-sections used we observe that the multi-component models exhibit a power-law behavior. The scaling of the relic abundance with DM mass is summarised in Table~\ref{tab:summary}. The power laws are a good fit for the heavier components in the set-up and do not apply to the lightest component.\footnote{In order for the expressions to be dimensionally correct, the scalings go as $(m_i/m_N)^k$, where $k$ is the exponent shown in the table and $m_N$ the lightest mass. We do not show this explicitly in the table.}

\begin{table}[!htb]
\centering
\begin{tabular}{|c|c||c|c|c|}
\hline
&Cross-section& \multicolumn{2}{c|}{\textbf{Symmetric}} & {\textbf{Partially-Asymmetric}}\\
\hline \hline
Model &$\mathbf{\sigma}$ & Ann. & Ann.$+$Conv. & Ann./ Ann.$+$Conv.\\
\hline
\textbf{Mass-independent} &$\neq f(m_i)$ & $\sim m^0$ & $\sim m^{-1}$ & $\sim \eta\, m^{[0,1]\,\ast}$\\
\hline
\textbf{Heavy mediators} &$\sim m_i^2/ \Lambda^4$ & $\sim m^{-2}$ & $\sim m^{-[3,2]}$ & $\sim \eta\, m^{[-2, 1]\,\ast}$\\
\hline
\textbf{Light mediators} &$\sim 1/m_i^2$ & $\sim m^{2}$ & $\sim m^{[0, 2]}$ & $\sim \eta\, m^{[1,2]}$\\
\hline
\end{tabular}
\caption{\label{tab:summary} Scaling of the relic abundance with the DM mass for the heavier components in a multi-component DM with symmetric or asymmetric components, with and without conversions, for different types of dependence of the cross sections on the DM mass. The left/right exponents indicate the extreme limits that are obtained for strong/weak conversions (and annihilations) in the symmetric (partially-asymmetric) case. In the cases with an $\ast$, left/right exponents indicate weak/strong interactions.}
\end{table}

We show the case of treating all cross sections as mass-independent ($\sigma \neq f(m_i)$), as well as the cases of light and heavy mediators. In the mass-independent symmetric case without conversions, all of the particles contribute with a similar abundance to the total one. Similarly, when the cross sections are inversely proportional to the DM mass, e.g. for the light mediator case, strong conversions may lead to several stable components with similar abundances in the symmetric case, e.g. the relic abundance becomes independent of the DM mass. On the other hand, in models with heavy mediators, we expect the lighter components of the dark sector to dominate the relic abundance in presence of significant conversions. 

For the partially-asymmetric case, if the annihilations and conversions are strong, the heavier components are highly asymmetric and the abundance depends linearly on mass, whereas if both the processes are weak, we recover the dependence found in the symmetric case with weak conversions. Therefore, depending on the values of interaction cross-sections, it may be possible to have several partially-asymmetric particles with similar abundances in the heavy mediator case or in the mass-independent case with weak annihilations and conversions.

\subsection{Abundance of the lightest state \label{sec:lightest}}

The power laws provided above are a good fit to describe the behavior of heavier components that undergo conversions. However, this is not true for the lightest component that is being produced by all the heavier components. In this section, we comment on an interesting effect that the lightest component undergoes in presence of conversions, which to our best of knowledge has not been discussed elsewhere. 

\begin{figure}[!htb]
\centering
\includegraphics[width=0.49\linewidth]{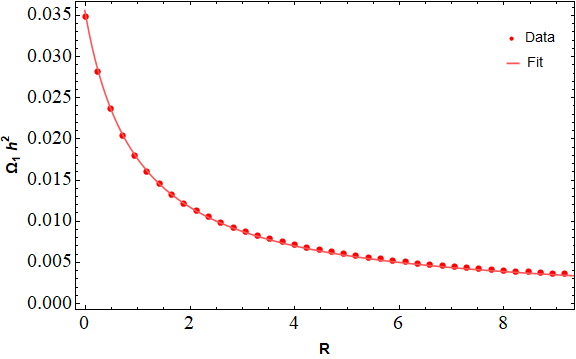}
\includegraphics[width=0.49\linewidth]{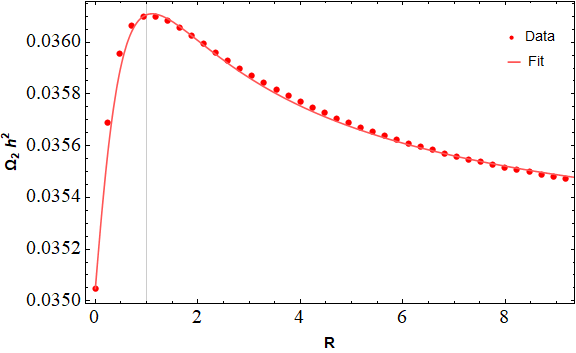}
\caption{Abundance of the heaviest (\emph{left panel}) and the lightest component (\emph{right panel}) versus $R=\con/\annone$ for the case of mass-independent cross-sections with $N = 2$, $\annone =\anntwo = 5.85 \times 10^{-9} {\rm~GeV}^{-2}$, $m_1 = 40$ GeV and $m_2 = 20$ GeV. The data is fitted with Eq. \eqref{peakmodelized}, with $\alpha = 9.38 \times 10^{-5}$, $\beta = 1.65$ and $\gamma = 0.77$. The peak close to $R \simeq 1$.} \label{fig:mipeak}
\end{figure}

It turns out to be useful to define $R = \con/\annone$ as the ratio of the conversion and the annihilation thermally-averaged cross-sections. In Fig.~\ref{fig:mipeak}, we plot the abundance of the heaviest and the lightest component versus $R$, for $N=2$, in the mass-independent case. In general, one would expect the abundance of the lightest (heaviest) component to be directly (inversely) proportional to the strength of conversions (i.e. from annihilations of the heavy states into it). While the abundance of heaviest does decrease with conversions, for the lightest component, we find that the abundance does not scale directly with the conversion rate. We observe the presence of a peak in the abundance when $R \simeq 1$, after which the abundance drops but is always (slightly) larger than the case without conversions ($R=0$). We observe that the peak appears at $R \simeq 1$, a phenomenon we term as `annihilation-conversion resonance'. The presence of this peak is inherent to conversions and can be attributed to the interplay of freeze-out dynamics for all the components involved. A word of caution, however, is in order: for $R \gg 1$, the computation of the relic abundance becomes more complicated, with the possible presence of significant reheating in the dark sector~\cite{Berlin:2016gtr}.

A suitable fit to the profile of the peak of the lighter component in Fig.~\ref{fig:mipeak} (right panel) is given by
\begin{equation}
   \Omega_lh^2(R) = \Omega_lh^2(0) + \frac{\alpha R}{R^{\beta} + \gamma}\,,
    \label{peakmodelized}
\end{equation}
where $\alpha$ and $\gamma$ are constants to be determined from the fit, and $\Omega_l h^2 (R)$ [$\Omega_lh^2(0)$] is the abundance of the lightest component in the case of conversions [no conversions]. For the heavier components (Fig.~\ref{fig:mipeak}, left panel), the fit goes as $\alpha^\prime/(1+R)$, where $\alpha^\prime$ is a constant. 

\begin{figure}[!htb]
	\centering
	\includegraphics[width=0.49\linewidth]{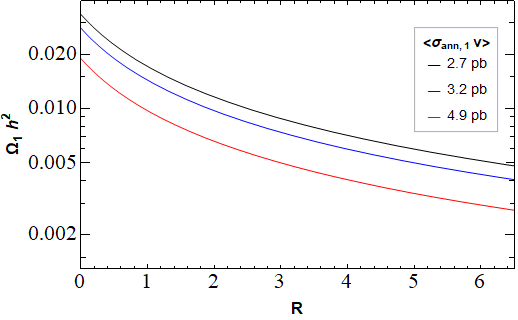}
	\includegraphics[width=.49\linewidth]{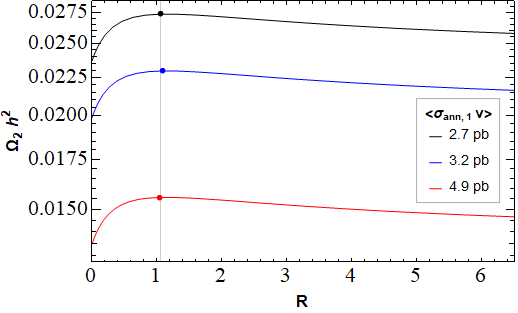}
	\caption{Abundance of the heaviest (\emph{left panel}) and the lightest component (\emph{right panel}) for different values of annihilation cross-sections in \emph{the light mediator case} for $N = 2$, $m_1 = 1200 {\rm~GeV}$ and $m_2= 1000 {\rm~GeV}$.}
	\label{fig:lightmed}
\end{figure}

Such peaks in the lightest component abundance can be seen in all the scenarios discussed above, whether the DM is symmetric/partially asymmetric or heavier/lighter than the mediator involved, for $N \geq 2$, as long as the mass splitting between the DM components fulfills $\Delta_{ij} \geq 0.05$. Note that the mass splitting $\Delta_{ij}$ of both DM species plays a central role in the above discussion as a consequence of Eq.~\eqref{eq:miconv}, since for small  $\Delta_{ij}$ no peak in the abundance is observed. In Fig.~\ref{fig:lightmed}, we show the case of mass-dependent cross-sections, using the parameterization of Eqs.~\eqref{eq:miannv} and \eqref{eq:miconv}. We consider the light mediator case with different values of $a$ (annihilations) and $c$ (conversions).

\section{Phenomenological Implications \label{sec:pheno}}

In general, the phenomenological implications are model dependent, and mostly driven by the interaction strength with the SM (annihilations), and to a lesser extent by conversions. Assuming similar values for the different components, the following generic features are expected.

In multi-component DM scenarios, if there are no conversions, typically the component with the smallest annihilation cross section dominates the abundance. Given that indirect detection (ID) signals are proportional to $n^2 \braket{\sigma_{\rm ann,i}v} \propto 1/(m^2 \braket{\sigma_{\rm ann,i} v})$, for equal mass-independent annihilation cross sections, we expect the lightest component to dominate. In presence of conversions, the abundance of the heavier components gets reduced and thus the ID signals should be suppressed compared to the case with no conversions. For example, in 2DM case, the ID rates for the heavier component $\chi_1$ and the lighter component $\chi_2$ are
\begin{align}
{R_{\rm sym,1}^{\rm ID}}&=\frac{1}{m_1^2}\frac{\annone}{(\annone+\con)^2}\,,\nonumber\\
{R_{\rm sym,2}^{\rm ID}}&=\frac{1}{m_2^2}\frac{1}{\anntwo}\,.
\end{align}
The scaling of indirect detection rates with DM mass for the different scenarios are shown in Table \ref{tab:ID}. We observe that in the light mediator case with only annihilations, or in presence of weak conversions, the ID rates of the different DM components are of similar order.

\begin{table}[!htb]
\centering
\begin{tabular}{|c|c||c|c|}
\hline
&Cross-section & \multicolumn{2}{c|}{\textbf{Indirect Detection}} \\
\hline \hline
Model &$\mathbf{\sigma}$ & Only Ann. & Ann.$+$Conv. \\
\hline
\textbf{Mass-independent} &$\neq f(m_i)$ & $\sim m^{-2}$ & $\sim m^{-4}$ \\
\hline
\textbf{Heavy mediators} &$\sim m_i^2/ \Lambda^4$ & $\sim m^{-4}$ & $\sim m^{-[6,4]}$ \\
\hline
\textbf{Light mediators} &$\sim 1/m_i^2$ & $\sim m^{0}$ & $\sim m^{[-4, 0]}$\\
\hline
\end{tabular}
\caption{\label{tab:ID} Indirect detection rates proportional to $n^2 \langle \sigma\,v \rangle $ for heavier components in a multi-component DM with symmetric components, for different scenarios. The exponents (left/right) indicate the extreme limits that are obtained depending on the strength of conversions (strong/weak).}
\end{table}

In the partially-asymmetric case with 1DM, there is an exponential reduction of the asymmetry ratio (and therefore on the indirect signals) with the annihilation cross section compared to the vanilla symmetric case, as pointed in Ref.~\cite{Graesser:2011wi}. In the case of partially-asymmetric multi-component DM, we find an exponential dependence of  the asymmetry ratios on the conversion cross section for the heavier species, making them more asymmetric, whereas the lighter ones may remain mostly symmetric. Therefore, with strong conversions, the heavier species become even more asymmetric, e.g. $r_i \ll 1$, and their indirect detection signals are highly suppressed compared to a WIMP candidate of the same mass~\cite{Graesser:2011wi},
\begin{equation}\label{eq:asymid}
\frac{R_{\rm asym}^{\rm ID}}{R_{\rm sym}^{\rm ID}}=\frac{\sigma_{{\rm ann},\,h}}{\sigma_{\rm sym}}\times r_{h,\infty}\times \left(\frac{2}{1+r_{h,\infty}}\right)^2 \ll 1\,.
\end{equation}
The ID rates for lighter components are also suppressed but not as much as the heavier ones
\begin{equation}\label{eq:asymliid}
\frac{R_{\rm asym}^{\rm ID}}{R_{\rm sym}^{\rm ID}}=\frac{\sigma_{{\rm ann},\,l}}{\sigma_{\rm sym}}\times r_{l,\infty}\times \left(\frac{2}{1+r_{l,\infty}}\right)^2 \lesssim 1\,.
\end{equation}
On the other hand, if the conversions are weaker and the annihilation rates are similar to the thermal one, the symmetric part of the components may be similar, as already discussed above. 

One may wonder whether strong conversions may lead to a locally-enhanced population of the lighter states compared to the heavier ones in high DM-density environments, such as the galactic centre or dwarf galaxies, enhancing the previous rates. Schematically, assuming that the local energy density $\rho_{\rm local}$ in the considered body scales as the global one $\Omega_{\rm global}$, as is the case for cold DM ~\cite{Bertone:2010rv,Blennow:2015gta}, we may have
\begin{equation}
\frac{\rho_{\rm local}^{\rm light}}{\rho_{\rm local}}>\frac{\Omega^{\rm light}_{\rm global}}{\Omega_{\rm global}}\,,\quad \frac{\rho_{\rm local}^{\rm heavy}}{\rho_{\rm local}}<\frac{\Omega^{\rm heavy}_{\rm global}}{\Omega_{\rm global}}\,.
\end{equation}
If the lighter states become more populated, their annihilation channels, say into SM states $\alpha$, will dominate. This would be interesting phenomenologically, given the stronger ID limits on light DM. For example, in the case of 2DM with masses $m_{1,2}$ ($m_1>m_2$) and number densities $n_{1,2}$, conversions will lead to an increase in the number density of the lighter component. One can estimate
\begin{equation}
\Delta n_2 \simeq 2\, n_{1,i}^2\, \con \, t\,,
\end{equation}
where $t$ is the relevant time scale from the formation of the local body, and $n_{1,i}$ is the initial local number density of the heavier component at the time where the body is formed. The change in the lighter species may be estimated as
\begin{align}
\Delta n_2 \simeq 10^{-9} {\rm~cm}^{-3} \left[\left(\frac{\mathcal{P}}{0.67}\right)\left(\frac{\rho_{\rm local}}{3 {\rm~ GeV}/{\rm cm}^{3}}\right) \left(\frac{10 {\rm~GeV}}{m_1}\right)\right]^2 \left(\frac{\con}{10^{-8}{\rm~GeV}^{-2}}\right)\left(\frac{t}{8 {\rm~Gy}}\right)\,,
\end{align}
where $\mathcal{P}\equiv \rho^1_{\rm local}/\rho_{\rm local}=\Omega_{\rm global}^1/\Omega_{\rm global}$ is the initial fraction of the energy density in the heavy state. Therefore, the  current local number densities of the states become
\begin{equation}
n_{1,f} \simeq n_{1,i} - \Delta n_2\,,\quad n_{2,f} \simeq n_{2,i} + \Delta n_2\,.
\end{equation}
Though in this case the change is negligible, it could be significant for local bodies with much higher dark matter densities, for lighter DM, and for larger conversions. A detailed analysis of the evolution of those systems would be needed in that case.

Multi-component DM could also give signatures in direct detection, specially kinks in the energy spectra of averaged rates~\cite{Profumo:2009tb,Herrero-Garcia:2017vrl}, and partial cancellations and a non-sinusoidal behavior in time-dependent rates~\cite{Herrero-Garcia:2018mky} (being able to reproduce DAMA Phase 2 results). The implications on direct detection rate for reproducing the relic abundance in thermal-freeze-out and asymmetric scenarios have been discussed in Ref.~\cite{Herrero-Garcia:2018qnz}. In general, the presence of large conversions allows to reproduce the relic abundance with smaller DM-SM couplings, hence relaxing the strong bounds coming from DD null-results.

Astrophysical signals from capture in the sun and other celestial bodies (e.g. neutron stars) can constrain the set-up. For example, the asymmetric component may yield constraints from creating black-holes, etc. The symmetric part may yield the typical signals, for instance annihilation of DM components into neutrinos, and/or anomalous heating of the celestial body. The conversions $\chi_1\chi_1 \rightarrow \chi_2 \chi_2$ with $m_1 \simeq m_2$ (inelastic scattering) can yield a larger population of the lowest mass state in the Sun (assuming thermalisation is reached) \cite{Blennow:2015hzp}. The opposite reaction $\chi_2\chi_2 \rightarrow \chi_1 \chi_1$ is kinematically forbidden in typical bodies for non-negligible mass splittings.

Similarly, the conversions $\chi_1\chi_1 \rightarrow \chi_2 \chi_2$ with $m_1 \simeq m_2$ (inelastic scattering) can explain small-scale structure problems in dwarfs  if large enough (also bounds from clusters)~\cite{Blennow:2016gde} and again yield a larger population of the lowest mass state. The opposite reaction $\chi_2\chi_2 \rightarrow \chi_1 \chi_1$ is typically kinematically forbidden also in clusters for non-negligible mass splittings.

At colliders, even if strong conversions make the DM to be mainly composed of the lighter state, the heavier states, if kinematically accessible, may also be produced. For LHC studies of multi-component scenarios see for example Ref.~\cite{Giudice:2011ib,Bhattacharya:2018cgx}.

\section{Conclusions\label{sec:summary}}

Guided by the visible sector, one may argue that the most natural dark sector is multi-component and partially-asymmetric. In this work, we have studied the relic abundance of several stable particles in the dark sector with symmetric and asymmetric components, with special emphasis on the role that conversions have in the final relic abundance. We have found that the total abundance gets reduced by conversions in all cases, generically helping to \emph{hide} the dark sector from experiment.

In many scenarios, we have found that several DM components may have similar abundances, contributing on equal footing to the relic abundance. These are the cases of: \emph{i)} symmetric or partially-asymmetric DM with mass-independent cross sections and small conversions, or light-mediators with strong conversions; \emph{ii)} heavy-mediator with weak conversions. We expect the dependence on the DM mass shown in Table \ref{tab:summary} to be valid for a large class of models, and to be rather universal as long as the masses and cross sections are at the weak scale.

In the symmetric case, for mass-independent cross sections, the abundance of heavier species is suppressed due to significant conversions and they are mostly asymmetric, while the lightest ones dominate the energy density. Furthermore, for large conversions and low masses, the local density of lighter states may be enhanced with respect to the global one in very high-density environments. These effects specially affect ID signals, with lighter states being more populated and yielding the largest SM final state products, while heavier ones remain mostly asymmetric and less populated.

We conclude that, in multi-component scenarios, it is natural to have symmetric and asymmetric components in presence of conversions. These can help to explain the absence of any positive signal up to now, and furthermore provide distinctive experimental signatures in current and future direct and indirect detection experiments, as well as at colliders, where cosmologically subdominant components may be produced.

\vspace{1cm}
\textbf{Acknowledgments} 

This work is supported by the MICIN/ AEI (10.13039/501100011033) grants PID2020-113334GB-I00 and PID2020-113644GB-I00. JHG and DV are supported by the “Generalitat Valenciana” through the GenT Excellence Program (CIDEGENT/2020/020). AB is supported by the \emph{JAE Intro Program, CSIC}.

\newpage
\appendix

\section{Model Example \label{sec:model}}

Here we provide an example model to realize multi-component DM and demonstrate the behavior of the cross section with DM mass, in line with the discussion in Section~\ref{sec:numerics}. One of the simplest models where this can be seen is the scalar singlet DM model, where the DM couples to the standard model via Higgs portal \cite{McDonald:1993ex,Burgess:2000yq}. Most of the parameter space of the real scalar singlet model is ruled out, leading to DM masses either in the funnel region of Higgs resonance or $m_{\rm DM}>1\rm{~TeV}$~\cite{Cline:2013gha, Casas:2017jjg}. In the following, we use it as an illustration and extend it to $N$ components. The parameter space for the two component model has been studied in detail in Ref.~\cite{Bhattacharya:2016ysw}.

Consider the standard model extended by $N$ scalar singlets $\phi_i$, all of them stabilised by $Z_2 \times Z_2^{'} \ldots \times Z_2^{'N}$ symmetry and coupling to the SM via a Higgs portal coupling $\l_i$. The part of the Lagrangian concerning DM interactions is given by 
\begin{align}
{\mathcal L}_{\rm DM-SM}&= \sum_{i=1}^N \bigg(\frac{1}{2}(\partial_\mu\,\phi_i)^2 -\frac{1}{2}m_{0_i}^2 \phi_i^2 -\frac{1}{4!}\lambda_{s_i} \phi_i^4 -\frac{1}{2}\lambda_{i}\phi_i^2 H^\dagger H -\frac{1}{4}\sum_{j>i} \lambda_{ij} \phi_i^2 \phi_j^2 \bigg)\,,
\end{align}
where $\l_{s_i}$ is the quartic self coupling, $m_{0_i}$ is the DM bare mass, $\l_{ij}$ is the DM-DM coupling and $H$ is the SM Higgs doublet, which acquires a vacuum expectation value after electroweak symmetry breaking and is parameterised by $H=(0,(\upsilon+h)/ \sqrt{2})^T$, where $\upsilon =246 {\rm~GeV}$, while $\braket{\phi_i}=0$. Hence, the mass of the DM candidates are $m_i^2 = m_{0_i}^2 + \frac{\lambda_i \upsilon^2}{2}$ and we choose $m_i > m_j$ for $j >i$. The Higgs portal couplings $(\lambda_{i})$ opens annihilation channels for DM states, $\phi_i \phi_i \r ff, WW, ZZ, hh$. Furthermore, they also lead to conversions $\phi_i \phi_i \r \phi_j \phi_j$ along with the direct interactions $\lambda_{ij}$, which dominate for larger DM masses.

In the large DM mass regime (light mediator case), $m_i \gg m_h$, the annihilation cross section ($s$-channel) for the model can be roughly approximated as
\begin{equation}
\braket{\sigma_{\rm ann,\,i}\,v} \approx \frac{\lambda_i^2}{16\pi\,m_i^2}\quad {\rm for~} m_i \gg m_h\,.
\end{equation}
According to our adopted convention, this implies $a_i \equiv \l_i^2/16\pi$.

For conversions, the interaction rate is given by
\begin{align}
\braket{\sigma_{\rm conv,\,ij}\,v} &= \frac{\sqrt{s-4 m_j^2}}{8\pi s\sqrt{s}}\bigg[
\frac{\upsilon^4 \l_i^2 \l_j^2}{(s-m_h^2)^2 +m_h^2 \Gamma_h^2}+\frac{2(s-m_h^2)\upsilon^2 \l_i \l_j \l_{ij}}{(s-m_h^2)^2 +m_h^2 \Gamma_h^2}+\l_{ij}^2 \bigg]\,.
\end{align}
Taking the threshold value $s = 4m_i^2$, in the heavy mass regime, the direct interaction term $\l_{ij}$ dominates over $\l_{i,j}$, and we can approximate
\begin{equation}
\braket{\sigma_{\rm conv,\,ij}\,v} \approx \frac{\l_{ij}^2 (m_i^2 -m_j^2)^{1/2}}{32\pi m_i^3}\,.
\end{equation}
Therefore, $c_{ij} \equiv \l_{ij}^2/32\pi$.

In the low DM mass regime (heavy mediator case), the annihilation and conversion cross sections can be approximated by
\begin{align}
\braket{\sigma_{\rm ann,\,i}\,v} &\approx \frac{\lambda_i^2}{8\pi}\,\frac{ m_i^2}{m_h^4}\quad {\rm for~} m_i \ll m_h\,,\nonumber\\
\braket{\sigma_{\rm conv,\,ij}\,v} &\approx \frac{\l_j^2}{8\pi\,m_h^4}\,m_i (m_i^2 - m_j^2)^{1/2}\,.
\end{align}
The approximation for the conversion cross section above holds in the case where the DM mass is given by $m_i^2 \sim \l_i \upsilon^2$. Thus, $b_i \equiv \l_i^2/8\pi$ and $d_{ij} \equiv \l_j^2/8\pi$ in this case.

\bibliographystyle{JHEP}
\bibliography{JHEPv1}

\end{document}